\documentclass[12pt]{article}
\usepackage[margin=1.2in]{geometry}
\usepackage[onehalfspacing]{setspace}
\usepackage{amsmath}
\usepackage{amsfonts}
\usepackage{amssymb}
\usepackage{booktabs}
\usepackage{physics}
\usepackage[normalem]{ulem}
\usepackage[bottom]{footmisc}
\usepackage{mathtools,amsthm,amssymb,physics}
\usepackage{smartdiagram}
\usepackage{multirow}
\usepackage{diagbox}
\usepackage{tcolorbox}
\usepackage{mdframed}
\usepackage{mathrsfs}
\usepackage{natbib}
\usepackage{subcaption}
\usepackage[title,titletoc,toc]{appendix}
\usepackage[font=small,labelfont=bf,justification=justified]{caption}

 \usepackage{pgfplots}
\pgfplotsset{width=12cm,compat=1.9}
\usepackage{pst-func}
\psset{unit=2cm}
        \usepackage{hyperref}

        \hypersetup{colorlinks=true,urlcolor=blue,citecolor=blue,linkcolor=red}

        \usepackage{caption}
	\usepackage{comment}
	\usepackage{setspace}
 	
	\usepackage{epsfig}
	
	\usepackage{graphics}
	
	\usepackage{lscape}

	\usepackage[english]{babel}
	
	\usepackage{color}
	
	\DeclareMathOperator*{\argmax}{arg\,max}

        \theoremstyle{plain}\newtheorem {corollary}{Corollary}

        \theoremstyle{plain}\newtheorem {proposition}{Proposition}

        \theoremstyle{plain}\newtheorem {thm}{Theorem}[section]
         \theoremstyle{plain}\newtheorem {prop}{Propostion}[section]
        \theoremstyle{plain}
        \theoremstyle{plain}\newtheorem{lem}{Lemma}

        \theoremstyle{definition}

        \theoremstyle{definition}\newtheorem{mydef}{Definition}

        \theoremstyle{remark} \newtheorem{example}{Example}

        \theoremstyle{remark} 

        \theoremstyle{remark} \newtheorem*{rem}{Remark}
         \theoremstyle{remark} 
         
\begin{document}
\bibliographystyle{plainnat}

\title{\textbf{A Frequentist Approach to Revealed Preference Analysis}\thanks{%
The authors thank Herakles Polemarchakis, Costas Cavounidis, Robert Akerlof, Roy Allen, Matt Polisson, Victor Aguiar, Motty Perry, Daniele Condorelli, Ben Lockwood, Sharun Mukand, Carlo Perroni, and Ludovic Renou for helpful discussions, comments, and encouragement, as well as audiences at CRETA (Warwick) and the Bristol Econometric Study Group Conference 2025. Any remaining errors, as well as the opinions expressed herein, are our own.
}}
\author{Charles Gauthier\thanks{Universitat de Barcelona; \href{mailto:charles.gauthier@ub.edu}{\tt charles.gauthier@ub.edu}}\quad  Raghav Malhotra\thanks{University of Leicester; \href{mailto:r.malhotra@leicester.ac.uk} {\tt r.malhotra@leicester.ac.uk}} \quad Agust\'{i}n Troccoli Moretti\thanks{Universitat Pompeu Fabra and BSE; \href{mailto:agustin.troccoli@upf.edu}{\tt agustin.troccoli@upf.edu}}}

\date{February 7, 2026}
\maketitle

\begin{abstract}
\noindent
This paper develops a framework to study the statistical power of revealed-preference tests. With randomly sampled budgets and mild smoothness of demand, statistical learning implies that any model consistent with the data must approximate true choice behaviour. We interpret this result as follows: passing a revealed-preference test is informative only to the extent that the data are sufficiently rich to rule out economically meaningful departures from the maintained model. We make this precise by linking sample size and confidence to the magnitude of detectable departures, and by characterising how power rises with additional observations. Extending our approach beyond revealed-preference inequalities to smooth functional restrictions yields practical tests, even when exact revealed-preference tests are computationally infeasible. We also provide confidence intervals for smooth functionals of demand, including welfare effects. Simulations show that standard sample sizes can generate widely different power across models, contextualizing why some conditions ``rarely reject'' in practice.
\end{abstract}
  
        \bigskip
        {\bf Keywords}: revealed preference, finite data, preferences, hypothesis testing 

        {\bf JEL classification}: C12, C91, D0, D11, D12
\pagebreak

\section{Introduction}
Analysts often seek to verify whether a decision maker (DM) is consistent with a given model, using finite choice data, to ensure that policy predictions about behavior and welfare are reliable. A common approach is revealed preference (RP) analysis, which provides a set of exhaustive inequalities that are satisfied if and only if the DM's choices are consistent with a given class of preferences.\footnote{Following the terminology of \cite{varian1983non}, we refer to a set of inequalities as exhaustive if they are necessary and sufficient for the data set to pass the RP test.} RP tests typically assume that data consist of a finite number of non-stochastic observations. A subtle issue arises when these observations pass the RP test. Namely, the test cannot distinguish between a DM who is truly consistent with the model and one who is not, but for whom there were insufficient data to detect a deviation. 

The frequentist approach to addressing this problem in the context of hypothesis testing focuses on statistical power, the probability of rejecting a decision maker’s choices under the assumption that the DM does not adhere to the model. This approach requires constructing parsimonious alternative hypotheses. However, as noted by \cite{adams2015models}, \textit{``The difficulty is that there are many alternatives to rational choice models and no obvious benchmark.''} In other words, there is no straightforward way to assess the power of RP tests.\footnote{For example, \cite{bronars1987} proposes a power test based on the assumption that choices are uniformly random on the budget frontier.} This paper shows how results from statistical learning theory can be used to circumvent this limitation and enable power analysis.

To illustrate the problem, suppose an analyst observes a consumer making grocery purchases on ten different occasions, each time facing different prices. The consumer's choices satisfy all the revealed preference conditions—there are no apparent ``mistakes''. Should the analyst conclude that the consumer is rational? Not necessarily. Ten observations may simply be too few to detect irrationality, even if it is present. With only ten price--consumption pairs, many non-rational choice patterns can slip through undetected, much as a student who answers randomly on a ten-question multiple-choice exam might still pass by chance.\par 

This raises a natural question: how many observations $n$ are needed to detect with high probability a consumer who deviates from rationality by an amount $\varepsilon$? Standard revealed preference analysis cannot answer this question: it tells us whether observed choices could have come from a rational consumer, not whether the available data are sufficient to reliably detect deviations.

Our paper fills this gap using a simple idea. Suppose we observe a consumer's choices at $n$ randomly selected price points. Because demand functions are smooth (satisfying a Lipschitz condition), any function that fits these observations must be close to the consumer's actual choice function everywhere—not just at the observed points—once $n$ is sufficiently large. Consequently, if the true choice function is far from rational, so will any function consistent with the data, and the revealed preference test will reject rationality. If the consumer is rational, the test will never reject. By quantifying the number of observations needed as a function of the detection threshold $\varepsilon$, we show that explicit power guarantees are possible.

\par

Formally, we consider an economic model $g \in \mathcal{G}$, where $\mathcal{G}$ is a class of models satisfying a Lipschitz condition, which maps an exogenous variable (prices) into an endogenous variable (demand). The analyst observes exogenous prices $p_t$ and demand $x_t$ related by $x_t = g(p_t)$. Consider some subclass $\mathcal{F}\subset \mathcal{G}$ and define the ball of size $\varepsilon>0$ around $\mathcal{F}$ as
\[\mathcal{F}(\varepsilon):=\Big\{g\in \mathcal{G}: \inf_{g'\in \mathcal{F}}d(g,g')\leq \varepsilon\Big\}.\] 

\noindent
Suppose the analyst can sample $\{p_t\}_{t=1}^n$ independently at random. This generates a dataset of the form
\begin{equation}
    \label{eq: data set def}
    D_n=\{(p_k,g(p_k))\}_{k=1}^n.
\end{equation}

\noindent
Note that although $g$ is non-stochastic, the data are stochastic because prices are sampled at random. This aligns with the frequentist tradition in which the state of the world is deterministic and all stochasticity arises from the sampling procedure.

For given $\delta$, $\varepsilon  > 0$, we aim to find a computable function $n(\delta,\varepsilon)$ such that, whenever a dataset has more than $n(\delta,\varepsilon)$ observations, one can construct tests of size zero and power greater than $\delta$ to distinguish between the null hyposthesis $\mathbf{H}_0$ and the alternative $\mathbf{H}_1$:
\begin{equation}
    \label{eq: hypothesis test}
    \mathbf{H}_0: g \in \mathcal{F} \quad \text{vs.} \quad \mathbf{H}_1:g \in \mathcal{G}\setminus\mathcal{F}(\varepsilon).
\end{equation}
We show that if the class $\mathcal{G}$ is Lipschitz and the analyst has access to a revealed preference characterization of $\mathcal{F}$, such tests and function $n(\delta,\varepsilon)$ can be constructed for any choice of $\varepsilon$. Furthermore, we show that when the dataset size $n$ is fixed, one can compute a lower bound on the power $\delta(n,\varepsilon)$ and a lower bound on the closeness of preferences from the model $\varepsilon(n,\delta)$.

We then generalize our main results by studying cases where the analyst does not have access to RP characterizations, but rather some functional restriction $\mathcal{R}:\mathcal{G} \to \mathbb{R}$ of $\mathcal{F}$ such that:
\[\mathcal{R}(g)=0 \iff g \in \mathcal{F}.\] 

\noindent
Under some regularity conditions on $\mathcal{R}$, we can conduct the same analysis as above by constructing 
\[\mathcal{F}(\varepsilon)=\qty{g^\prime \in \mathcal{G}: \mathcal{R}(g)\leq \varepsilon},\]
and then forming the hypothesis test in the manner described in \eqref{eq: hypothesis test}. In that case, the size of the test is not zero, but some computable function of $\mathcal{G}$, $\mathcal{R}$, and $\mathcal{F}$.

Our method applies to preference classes where revealed preference characterizations are unknown or computationally infeasible. For Lipschitz demands, we find that the computational issues of several RP tests stem mostly from their knife-edge nature which have size zero. We demonstrate that this approach can be used to construct confidence intervals for smooth functionals of demand, such as changes in cost-of-living indices. Finally, we show how to integrate additional assumptions on the underlying preference classes and adaptive sampling to improve power and confidence intervals. 

We emphasize that the results in this paper are primarily existence results that establish what can be learned from finite choice data, rather than providing readily applicable procedures. Our theorems characterize the sample sizes sufficient to achieve desired power levels and the precision attainable for welfare inference, but operationalizing these bounds requires knowledge of quantities—such as the Lipschitz constant of demand and the sampling distribution of prices—that may be difficult to determine in practice. We view our contribution as providing the theoretical foundation for power analysis in revealed preference settings; developing practical implementations to estimate or bound these quantities from data is an important direction for future work.

Section \ref{sec:setup} describes our setup and notation. Section \ref{sec:results} shows how to construct our tests with fixed power, sample size, and distance from the model (Theorems \ref{thm: RP-PAC}--\ref{thm:min-detectable-separation}), followed by power simulations for RP tests. Section \ref{sec:app} extends our testing results to functional characterizations (Theorem \ref{thm:functional-test}) and shows how to conduct inference on smooth estimands such as welfare changes (Theorem \ref{thm:welfare-inference}). The section also provides functional characterizations of common classes of preferences, such as substitutes and complements, and of choice under uncertainty (Propositions \ref{prop:homothetic}--\ref{prop:Rbet}). Section \ref{sec:efficiency} studies extensions of our approach. Section \ref{sec:conclusion} concludes.

\subsection{Related Literature}

The RP approach tests the consistency of a finite dataset with utility maximization. The approach originated with \citet*{samuelson1938} and was refined by \citet*{afriat1967construction} and \cite{varian82}. In particular, \cite{afriat1967construction} established that the Generalized Axiom of Revealed Preference (GARP) is a necessary and sufficient condition for any finite dataset to be consistent with utility maximization. The appeal of GARP stems from the ease with which the rationality of a dataset can be verified through simple algebraic inequalities.



The idea of constructing an alternative against which to assess the power of revealed-preference (RP) tests dates back to \cite{becker1962irrational}, who considered consumers choosing randomly on their budget lines. Building on this, \cite{bronars1987} examined how often RP tests would detect violations when choices were generated uniformly at random. More recently, \cite{andreoni2013power} proposed a more realistic benchmark by drawing from the observed distribution of choices, thereby calibrating the alternative hypothesis to match empirical patterns.

Because a parsimonious alternative to fully rational choice is hard to specify, several papers assess predictive power without positing an explicit behavioural model. \cite{andreoni2013power} define the Afriat Power Index for datasets with no revealed-preference violations. Reporting the smallest budget perturbation needed to induce a violation—small adjustments indicate a sensitive test, large adjustments a weak one. A related approach, based on \cite{selten1991properties} and applied by \cite{beattycrawford11}, considers a method which uses the set of all theory-consistent behaviors with respect to all possible behaviors.

Since those seminal contributions, RP theory has explored various extensions, including functional-form restrictions and intertemporal models. Significant contributions include \citet*{kubler2014asset} for expected utility with objective (known) probabilities and \citet*{echeniquesaito15} for subjective probabilities. A similar problem for ``translation-invariant'' preferences is considered in \citet*{chambersetal16}. \citet*{polisson2020rp} give a general method to construct RP inequalities that applies to several classes of preferences over risk and uncertainty.\footnote{For recent extensive reviews, see \cite{CrawfordEmpirical2014}, \cite{EcheniqueRPreview2019}, and \cite{demuynck2019samuelson}.} We take the set of Lipschitz demands as the alternative, thus covering these subclasses and rationality itself.



The functional approach for testing decision-makers \citep{slutsky1915, antonelli71, DeatonMuell1980} relies on the entire demand function and assumes access to infinite data. We build on this literature by extending these tests to finite data. In this context, our work is related to \citet*{aguiar2018classifying}, who adapt the measure of bounded rationality from \citet*{aguiar2017slutsky} to finite data. While their focus is on measuring and classifying bounded rationality, our approach provides tests for general restrictions under preference monotonicity, making it applicable to a broader range of datasets, including pairwise comparisons.

Papers on extrapolating fundamentals from finite data use regularity conditions similar to ours. \cite{chambers2021recovering} imposes a structural restriction on the space of preferences that is closely related to Lipschitz continuity, but it allows sample complexity to depend on the underlying (unknown) preference, which we rule out by construction. Related Lipschitz-type upper bounds for utilities are also used in \cite{chambers2023recovering} to deliver finite-sample guarantee results. Finally, \cite{chambers2025decision} shows that economic models that approximately satisfy axioms can be rationalized by utilities close to those that exactly rationalize the model, providing microfoundations for using our approach to test axiomatic decision models.



The RP literature focuses on exhaustive restrictions in finite datasets that characterize a model.\footnote{The only revealed preference test that is not a characterization which we are familiar with is for probabilistic sophistication \citep*{Epstein2000probsoph}.} In some instances, those tests can pose a significant computational challenge. Indeed, \citet*{echenique2014testing} and \citet*{Cherchyeinteger2015} show that testing for weak separability is NP-Hard. Accordingly, these tests rely on non-polynomial-time algorithms, such as mixed-integer programming \citep*{Cherchyeinteger2015,Hjertstrandmixint2020}. We show that the computational issues can be mitigated by permitting tests with nonzero size.

The deterministic RP approach has recently been extended to stochastic environments by \cite{KS2018} and \cite{AK2021}. In both cases, RP analysis is embedded within a random utility framework, although the latter treats measurement error as the source of randomness. By contrast, in our framework, all randomness arises from the sampling procedure rather than from the decision-maker. Consequently, our approach is closer in spirit to standard RP tests conducted on individual-level data, while still treating individual observations as random.\footnote{In a different direction, \cite{allen2024} propose a statistical consumer model who chooses a distribution of demands given prices. They show that standard RP conditions apply to average demands under a weak mean‑expenditure condition, so our method should extend to their framework as well.}

\section{Setup}\label{sec:setup}


We consider a DM whose choices are observed a finite number of times. Let $Y\subset \mathbb{R}^\ell_{\geq0}$ be a compact and convex consumption space, where $\ell$ denote the number of goods. A DM faces prices $p\in \mathcal{P}$ and income $I\in \mathcal{I}$, where $\mathcal{P}$ is a compact subset of $\mathbb{R}^\ell_{>0}$ and $\mathcal{I} = [a,b] \subset \mathbb{R}_{>0}$. We denote the budget corresponding to a tuple $(p,I)$ by
\begin{equation*}
    B(p,I)=\Big\{y\in Y: p\cdot y \leq I \Big\},
\end{equation*}

 \noindent
 and denote the set of all budget sets by  $\mathcal{D}$. The DM has a choice function $x(p,I)$ with $x(p,I)\in B(p,I)$. Let $\mathcal{G}$ denote the class of all choice functions $x:\mathcal{P}\times\mathcal{I}\to Y$ that satisfy the following Lipschitz property: there exists a constant $L>0$, known to the analyst, such that for all $(p,I),(p',I')\in\mathcal{P}\times\mathcal{I}$,
\[
\|x(p,I)-x(p',I')\|\le L\, d\!\left((p,I)^{\top},(p',I')^{\top} \right),
\]
where $\|\cdot\|$ is the Euclidean norm and $d$ is the Euclidean distance.

 In several applications, we restrict attention to choice functions arising from utility maximisation.
A preference relation $\succeq\,\subseteq Y\times Y$ is \emph{rational} if it is complete and transitive. We further assume that preferences are continuous, strictly monotone and strictly convex.\footnote{
Strict monotonicity means that for any $x,y\in Y$, if $x_i\ge y_i$ for all $i=1,\ldots,\ell$ and $x\neq y$, then $x\succ y$.
Strict convexity means that for any $x\neq y$ in $Y$ and any $\alpha\in(0,1)$, $x\succeq y$ implies $\alpha x+(1-\alpha)y\succ y$.
}
Given a rational preference $\succeq$, let $u_\succeq:Y\rightarrow \mathbb{R}$ be a (continuous) utility representation and define the associated Marshallian demand by
\[
x_\succeq(p,I)\;:=\;\argmax_{y\in B(p,I)} u_\succeq(y).
\]
Under strict convexity this maximiser is unique, so $x_\succeq$ is single-valued. Accordingly, we say that a choice function $x\in\mathcal{G}$ is an \emph{admissible rational demand} if there exists a rational preference $\succeq$ such that $x = x_\succeq$. 
In what follows, we use the term \emph{choice function} for an arbitrary element of $\mathcal{G}$, and the term \emph{demand function} for an admissible rational demand.

A dataset consists of budget-choice pairs $D_n = \{(B_k, x_k)\}_{k=1}^n$. A choice function $x$ \textit{rationalizes} a dataset $D_n$ if it exactly reproduces the observed choices $x_k = x(B_k)$ for all $k=1,\ldots,n$.

Let $g\in \mathcal{G}$ denote the true choice function, which we may refer to as the underlying model. The analyst draws $n$ price--income pairs independently at random from some distribution $\mu \in \Delta(\mathcal{P} \times \mathcal{I})$, generating a dataset
\begin{equation*}
    D_n(x^*) \;=\; \big\{(B_k,\, g(B_k))\big\}_{k=1}^n.
\end{equation*}

\begin{figure}[h!]
\centering
\includegraphics[scale=1]{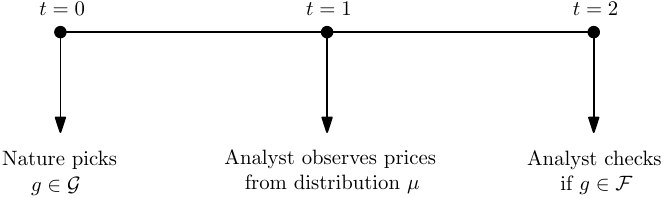}
\caption{Depiction of the timing in our framework.}
\label{f: timeline new approach}
\end{figure}

\noindent
Let $\mu^n$ denote the product measure on $(\mathcal{P}\times \mathcal{I})^n$.\footnote{Formally, if $\mu \in \Delta(\mathcal{P}\times\mathcal{I})$ is the sampling distribution of price--income pairs, then $\mu^n$ is the product measure on $(\mathcal{P}\times \mathcal{I})^n$, defined for any measurable set $A \subseteq (\mathcal{P}\times \mathcal{I})^n$ as 
\[
    \mu^n(A) \;=\; \int \mathbf{1}_A\big((p_1,I_1),\ldots,(p_n,I_n)\big) 
    \,\mathrm{d}\mu(p_1,I_1)\cdots \mathrm{d}\mu(p_n,I_n).
\]} 
Since $g\in\mathcal{G}$ is fixed, $\mu^n$ induces a distribution $\mu^n_g$ on datasets of size $n$ by mapping each sampled sequence  $\big((p_k,I_k)\big)_{k=1}^n$ to its corresponding budget-choice sequence $\big(B_k,g(B_k)\big)_{k=1}^n$. By treating the dataset as random draws from a fixed model, we can study the size and power of tests over repeated sampling. The timing of this conceptual framework is illustrated in Figure~\ref{f: timeline new approach}.



Let $\mathbf{H}_0$, $\mathbf{H}_1 \subset \mathcal{G}$ denote 
the null and alternative hypotheses. We have $g \in \mathbf{H}_0$ if the null hypothesis holds and $g \in \mathbf{H}_1$ if the alternative hypothesis holds. A \textit{test} is a measurable function $\mathcal{T}: \mathcal{D}\to \{0,1\}$ that takes as input a dataset and returns whether the null hypothesis should be rejected, 
where by convention $\mathcal{T}(D)=1$ means the null is rejected and $\mathcal{T}(D)=0$ means the null is not rejected.
\begin{mydef}[Size and Power]
\label{def:sizepower}
Let $\mathcal{T}$ be a test and $\mathcal T^{-1}(1)$ denote the set of datasets rejected by $\mathcal T$.
\begin{enumerate}
    \item[\textit{(i)}] The \textit{size} of $\mathcal{T}$ is the largest probability of incorrectly rejecting when the null is true:
    \[
        Size(n) \;=\; \sup_{x \in \mathbf{H}_0}\mu^n_{x}\!\left(\mathcal{T}^{-1}(1)\right).
    \]
    \item[\textit{(ii)}]  The \textit{power} of $\mathcal{T}$ is the probability of correctly rejecting the null when the alternative is true: 
    \[
        Power(n;g) \;=\; \mu^n_{g}\!\left(\mathcal{T}^{-1}(1)\right) \quad \text{for a given } g \in \mathbf{H}_1.
    \]
\end{enumerate}
\end{mydef}

The distinction between size and power lies in the domain over which probabilities are evaluated: size is defined uniformly over all models in the null, taking the supremum over $\mathbf{H}_0$, while power is evaluated at a particular alternative $g \in \mathbf{H}_1$. For brevity, we will henceforth omit the explicit dependence of $\mu^n_g$ on $g$ whenever there is no risk of confusion, and simply write $\mu^n$.

\begin{mydef}[RP Characterization]
\label{def:RPcharacterization}
Let $\mathcal{F}\subset \mathcal{G}$ be a class of preferences. An \textit{RP characterization} $\mathcal{T}^*$ of $\mathcal{F}$ is a test with the following property:  
\begin{enumerate}
    \item[\textit{(i)}] $\mathcal T^*$ never rejects a dataset rationalisable by some $x \in \mathcal F$ (size zero). 
    \item[\textit{(ii)}] For any other test $\mathcal T$ satisfying (i), if $\mathcal T$ rejects a dataset $D_n$, then $\mathcal T^*$ also rejects $D_n$ (maximal power).
\end{enumerate}
\end{mydef}

In words, an RP characterization never rejects a dataset consistent with $\mathcal{F}$, and has maximal power among all such size-zero tests.\par

\section{Finite-Sample Theory} \label{sec:results}
\label{sec: results}


This section develops finite–sample power bounds that connect the revealed preference framework to the statistical theory of PAC learning. We proceed in three steps. First, we determine the required sample size to achieve a specified power and accuracy level, which may inform experimental design. Second, for a given sample size and accuracy, we derive the attainable power level, helping to gauge the effectiveness of RP tests. Finally, we characterize the minimal separation from $\mathcal{G}$ that is detectable given a sample size and desired power.

\subsection{Preference Learnability}

This subsection introduces the notion of PAC learnability and states the main result we use as a building block for our procedure. First, define the generalization error between two choice functions $x$, $x':\mathcal{P}\times \mathcal{I} \to  Y$ as: 
\begin{equation*}
\mathsf{erf}_\mu(x,x')=\int_{\mathcal{P}\times \mathcal{I}} \norm{x(p,I)-x'(p,I)}\, \dd\mu,
\end{equation*}
where $\norm{\cdot}$ denotes the Euclidean norm. The next definition introduces the concept of PAC learnability.

\begin{mydef}[PAC Learnability]
\label{def:PAC Learnability}
A class of demand functions $\mathcal{G}$ is \textit{probably approximately correct} (PAC) learnable by a 
hypothesis class $\mathcal{H}$ if the following holds. For any $\varepsilon$, $\delta>0$, any choice function 
$x \in \mathcal{G}$, and any distribution $\mu \in \Delta(\mathcal{P}\times\mathcal{I})$, there exists an algorithm $L$ such that, given a sample of size
\[
m_L = m_L(\varepsilon,\delta)
    = \mathrm{Poly}\!\left(\tfrac{1}{\varepsilon},\,\tfrac{1}{\delta}\right),
\]
the algorithm returns a hypothesis $h \in \mathcal{H}$ satisfying $\mathsf{erf}_\mu(x,h) < \varepsilon$ with probability at least $1-\delta$.\footnote{We write $\mathrm{Poly}(1/\varepsilon,1/\delta)$ to indicate that the required sample size is bounded by a polynomial in $1/\varepsilon$ and $1/\delta$.}
\end{mydef}

\smallskip

In words, $\mathcal{G}$ is PAC learnable by $\mathcal{H}$ if there exists an algorithm that can approximate every demand function in $\mathcal{G}$ by some hypothesis $h \in \mathcal{H}$ within error $\varepsilon$, with probability at least $1-\delta$. Here, a hypothesis $h$ corresponds to a candidate demand function approximating the true choice function. While in the general learning-theory literature the hypothesis class $\mathcal H$ may differ from the target class $\mathcal G$, we restrict our attention to the case $\mathcal H = \mathcal G$. We can now state a key result for our derivations.


\begin{lem}
\label{thm: beigmanvohraPACleanrability}
    Let $\mathcal{G}$ be the class of choice functions over the compact set $\mathcal{P}\times\mathcal{I}$ with Lipschitz constant $L$. Then $\mathcal{G}$ is learnable by hypothesis class $\mathcal{H} = \mathcal{G}$ with sample complexity:
    \[ m_L(\varepsilon, \delta)=O\left(\frac{1}{\varepsilon^2} \left( \ln^2 \left( \frac{1}{\varepsilon} \right) \cdot \left(\frac{L}{\varepsilon}\right)^{\ell+1}+ \ln \left( \frac{1}{\delta} \right) \right)\right).\footnote{If, in addition, the choice rules are homogeneous of degree zero in $(p,I)$ and $I$ is bounded away from zero, we may normalise by income and obtain the same bound with exponent $\ell$.}\]
\end{lem}
This lemma guarantees that, for any distribution $\mu$, the true demand function can be approximated within any pre-specified accuracy $\varepsilon$, with confidence at least $1-\delta$, from a sample size that grows only polynomially in $1/\varepsilon$ and $1/\delta$. Our proof follows from a simple extension of \citet{beigman2006learning}, who prove that Lipschitz demand functions are learnable under the above sample guarantees. However, their proof relies on a simple packing-number argument that does not use the fact that the hypothesis class is assumed to consist of demand functions. They also provide a constructive algorithm that achieves this bound, thereby delivering both sample efficiency and polynomial-time computation.\footnote{\citet*{kubler2020identification} show that when the sampling distribution $\mu$ is uniform over $\mathcal{P}$, the assumption of a known Lipschitz constant can be relaxed, but they do not provide an explicit procedure to compute the relevant constants.}

For each $(\varepsilon,\delta)$, let $n(\varepsilon,\delta)$ denote the \emph{minimal} sample size such that there exists \emph{some} learning algorithm (possibly depending on $(\varepsilon,\delta)$) that $(\varepsilon,\delta)$--learns $\mathcal G$ in the sense of Definition~\ref{def:PAC Learnability}. In particular, for any specific algorithm $L$ satisfying Definition~\ref{def:PAC Learnability} with sample size $m_L(\varepsilon,\delta)$, we have $n(\varepsilon,\delta)\le m_L(\varepsilon,\delta)$. Moreover, $n(\varepsilon,\delta)$ is weakly decreasing in $\varepsilon$ for each fixed $\delta$.

\subsection{Finite–Sample Power Bounds}
\label{ss: finite-sample}


Our first result presents the “forward” formulation that characterises the sample size required to ensure that an RP test has power at least $1-\delta$ against all alternatives that are $\varepsilon$–separated from the null class.
\begin{thm}
\label{thm: RP-PAC}
Let $\mathcal{F} \subset \mathcal{G}$ be a class of models-, and suppose the analyst has access to an RP characterization $\mathcal{T}_{\mathcal{F}}$ for $\mathcal{F}$. Then, for every $\varepsilon$, $\delta > 0$, if the sample size $n$ satisfies $n\geq n(\varepsilon,\delta)$,\footnote{In particular, by Lemma~\ref{thm: beigmanvohraPACleanrability} it suffices to take $n=m_L(\varepsilon,\delta)$.}
the test based on $\mathcal{T}_{\mathcal{F}}$ applied to a dataset of size $n$ has power at least $1-\delta$ for the hypotheses
\[
\mathbf{H}_0: x^* \in \mathcal{F}
\qquad\text{vs.}\qquad
\mathbf{H}_1: x^* \in \mathcal{G} \setminus \mathcal{F}(\varepsilon),
\]
where
\[
\mathcal{F}(\varepsilon) \;:=\; \left\{ x \in \mathcal{G} : \inf_{x' \in \mathcal{F}} \mathsf{erf}_\mu(x,x') \,\leq\, \varepsilon \right\}.
\]
\end{thm}

\begin{figure}[h!]
    \centering
    \includegraphics[width=0.75\linewidth]{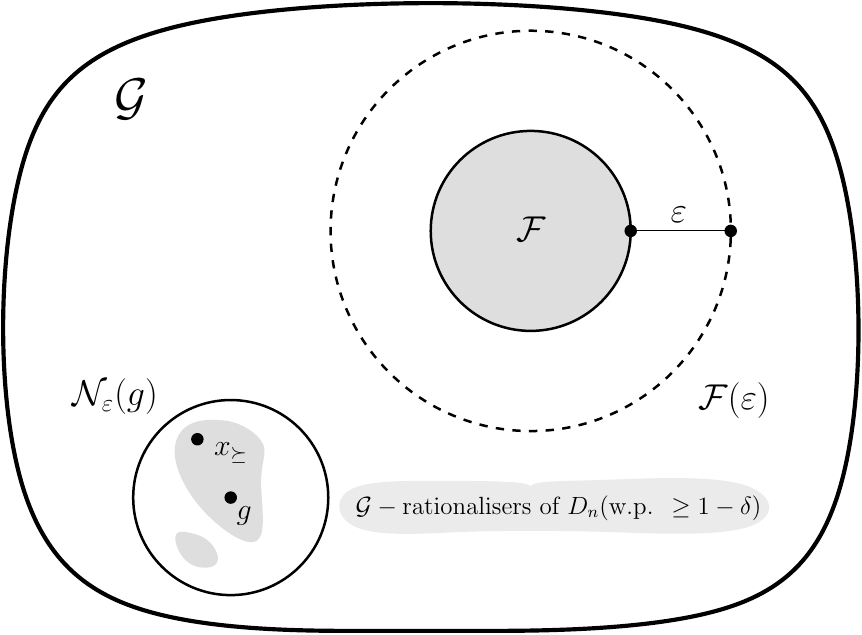}
    \caption{\textbf{Geometric sketch of the proof.} When $n\ge n(\varepsilon,\delta)$, with probability at least $1-\delta$, the $\mathcal{G}$-rationalisers of the dataset $D_n$ are contained in the $\varepsilon$-neighbourhood $\mathcal{N}_{\varepsilon}(g)$ of the true model (shaded region). Under the alternative $g\notin\mathcal{F}(\varepsilon)$, this neighbourhood is disjoint from $\mathcal{F}$. Hence no element of $\mathcal{F}$ can rationalise the data, and an RP characterisation rejects with probability at least $1-\delta$.}
    \label{fig:Proof Theorem 3.1}
\end{figure}

In words, this result guarantees that the test rejects any demand strictly more than $\varepsilon$ away from $\mathcal{F}$ with probability at least $1-\delta$ if the analyst collects any sample size of $n\geq n(\varepsilon,\delta)$ observations.\footnote{The distance $\mathsf{erf}_\mu(\cdot,\cdot)$ is computed using the same distribution $\mu$ that governs the sampling of price–income pairs. Thus, both the measure of approximation error and the test are based on the same underlying data distribution.} Here $n(\varepsilon,\delta)$ is the minimal sample complexity, and it suffices to take $n=m_L(\varepsilon,\delta)$. In particular, it is worth noting that Theorem \ref{thm: RP-PAC} allows to study power of rational preferences against an alternative Lipschitz demand. Indeed, one can simply take $\mathcal{F}$ to be the set of all choice functions that arise from utility maximisation, and the RP characterization for this class reduces to the standard Afriat inequalities. The geometric structure of the argument in Theorem \ref{thm: RP-PAC} is conveyed in Figure~\ref{fig:Proof Theorem 3.1}.



There, the set of $\mathcal{G}$-rationalising demands is represented by the neighbourhood $\mathcal{N}_{\varepsilon}(g)$. When $n \ge n(\varepsilon,\delta)$, with probability at least $1-\delta$, every $\mathcal{G}$-rationaliser of the observed dataset lies in $\mathcal{N}_{\varepsilon}(g)$. Under the alternative hypothesis, the true model satisfies $g\notin \mathcal{F}(\varepsilon)$, equivalently $\inf_{f\in\mathcal{F}}\mathsf{erf}_\mu(g,f)>\varepsilon$, so $\mathcal{N}_{\varepsilon}(g)\cap\mathcal{F}=\emptyset$. Consequently, with probability at least $1-\delta$, no element of $\mathcal{F}$ can rationalise the observed dataset. Since an RP characterisation rejects precisely when no element of $\mathcal{F}$ rationalises the data, it follows that the test rejects under the alternative with probability at least $1-\delta$.

While Theorem~\ref{thm: RP-PAC} provided the forward bound, the next result states the corresponding inverse bound: given a fixed sample size $n$ and tolerance $\varepsilon$, it determines the confidence level $1-\delta(\varepsilon,n)$ that can be guaranteed. The difference is only in which parameters are treated as primitive, so the proofs mirror each other. To avoid trivialities, for fixed $\varepsilon>0$ and $n\in\mathbb{N}$ we assume that there exists some $\delta\in(0,1)$ such that $n(\varepsilon,\delta)\le n$.

\begin{thm}
\label{thm:power-from-fixed-n}
Suppose the minimal sample complexity of $(\varepsilon,\delta)$-learning $\mathcal{G}$ is denoted by 
$n(\varepsilon,\delta)$. For each $\varepsilon>0$ and $n\in\mathbb{N}$, define
\[
\delta(\varepsilon,n) \;:=\; \inf\{\delta\in(0,1) : n(\varepsilon,\delta)\leq n\}.
\]
Let $\mathcal{T}$ be an RP characterisation of $\mathcal{F}$. Then, for any alternative $g\in\mathcal{G}$ with
\[
\inf_{f\in\mathcal{F}} \mathsf{erf}_\mu(g,f)\ \ge\ \varepsilon,
\]
we have
\[
\mu^n_{g}\big(\,\mathcal{T}^{-1}(1)\,\big)\ \ge\ 1-\delta(\varepsilon,n).
\]
In words, at tolerance $\varepsilon$ and sample size $n$, the test $\mathcal{T}$ has power at least $1-\delta(\varepsilon,n)$ against all alternatives at distance at least $\varepsilon$ from $\mathcal{F}$.
\end{thm}

Algebraically, Theorem~\ref{thm:power-from-fixed-n} amounts to inverting the sample--complexity map: instead of choosing $n$ large enough to guarantee a target confidence level $1-\delta$, we fix $(\varepsilon,n)$ and identify the \emph{best} confidence level compatible with $n$, namely $1-\delta(\varepsilon,n)$, where $\delta(\varepsilon,n)$ is the smallest $\delta\in(0,1)$ such that $n(\varepsilon,\delta)\le n$. Geometrically, the idea is the same as in Figure~\ref{fig:Proof Theorem 3.1}: with probability at least $1-\delta(\varepsilon,n)$, all $\mathcal{G}$-rationalisers of the dataset lie in the $\varepsilon$--ball around the true model, and if this ball is disjoint from $\mathcal{F}$ the test rejects.

Having stated the forward bound (fix $(\varepsilon,\delta)$, solve for $n$) and its inverse (fix $(\varepsilon,n)$, solve for $\delta(\varepsilon,n)$), it is natural to ask the complementary question: given a sample size $n$ and confidence level $1-\delta$, what separation $\varepsilon$ from $\mathcal{F}$ can the test reliably detect? The next result answers this by introducing the minimal detectable separation scale $\varepsilon(n,\delta)$. 
To avoid trivialities, we fix $n\in\mathbb{N}$ and $\delta\in(0,1)$ such that the set $\{\varepsilon>0:\ n(\varepsilon,\delta)\le n\}$ is nonempty. 

\begin{thm}
\label{thm:min-detectable-separation}
Let $n(\varepsilon,\delta)$ denote the minimal sample complexity of $(\varepsilon,\delta)$-learning $\mathcal{G}$. 
For $n\in\mathbb{N}$ and $\delta\in(0,1)$, define
\[
\varepsilon(n,\delta)\;:=\;\inf\{\varepsilon>0:\ n(\varepsilon,\delta)\le n\}.
\]
Let $\mathcal{T}$ be an RP characterisation of $\mathcal{F}$. Then, for every $\gamma>0$ and every alternative $g\in\mathcal{G}$ with
\[
\inf_{f\in\mathcal{F}} \mathsf{erf}_\mu(g,f)\ \ge\ \varepsilon(n,\delta)+\gamma,
\]
we have
\[
\mu^n_{g}\big(\,\mathcal{T}^{-1}(1)\,\big)\ \ge\ 1-\delta.
\]
In words, for any slack $\gamma>0$, at confidence level $1-\delta$ and sample size $n$, the test $\mathcal{T}$ has power at least $1-\delta$ against all alternatives at distance at least $\varepsilon(n,\delta)+\gamma$ from $\mathcal{F}$.
\end{thm}


The argument mirrors the proofs of Theorems~\ref{thm: RP-PAC} and~\ref{thm:power-from-fixed-n} and relies on the same geometry as in Figure~\ref{fig:Proof Theorem 3.1}. The key difference is that, for fixed $(n,\delta)$, the boundary tolerance $\varepsilon(n,\delta)$ is defined as an \emph{infimum} and need not be attained. Accordingly, the result is stated with an arbitrary slack $\gamma>0$.

Conceptually, Theorem~\ref{thm:min-detectable-separation} turns the sample--complexity map on its head: for a given sample size $n$ and confidence level $1-\delta$, it identifies a detectability frontier in terms of separation from $\mathcal{F}$. Any alternative whose distance from $\mathcal{F}$ exceeds $\varepsilon(n,\delta)$ by a strictly positive margin $\gamma$ is guaranteed to be detected with probability at least $1-\delta$ by any RP characterisation of $\mathcal{F}$. Thus, $\varepsilon(n,\delta)$ summarises the finite-sample resolution of the frequentist revealed-preference test: larger samples push the frontier down, while higher confidence shifts it up.

\begin{rem}
Notice that the power computation depends only on the sample complexity of PAC learning for 
$\mathcal{G}$. Appendix \ref{sec:efficiency} shows this can be exploited to improve power by presenting methods that accelerate learning rates.
\end{rem}

\begin{rem}
It is easy to show that if observed demands are subject to additive mean-zero measurement error with finite variance, then the class of learnable demand functions $\mathcal{G}$ remains learnable  \citep{bartlett1994fat}. Consequently, our previous results remain valid under additive mean-zero measurement error.
\end{rem}

\subsection{Power Analysis}\label{sec:implementation}

In what follows, we study the finite-sample power of revealed preference tests for static utility maximization (SUM), homotheticity (H), weak separability (WS), and expected utility (EU) against smooth violations of integrability. We consider a setup with four goods and, in the EU case, two equally probable states of the world. For the tests, we follow \citet{varian1983non} and \citet{green1986}. For weak separability, we only use necessary conditions since exact tests based on mixed-integer programming are NP-complete \citep{Cherchyeinteger2015}.

We consider simulated demands given by $y(p, I) = x(p,I) + \varepsilon f(p,I)$, where \(x(p,I)\) are Cobb-Douglas demands, \(\varepsilon>0\) controls the deviation from rationality, and \(f(p,I)\) is a perturbation that preserves homogeneity of degree zero and budget neutrality, but violates Slutsky symmetry.\footnote{Although the perturbed demands may be negative, revealed preference theory still applies \citep{deb2023}, so they are inconsequential for our simulations.} We normalize \(f\) such that the Frobenius norm of the skew-symmetric part of the Slutsky matrix equals one, ensuring comparability of \(\varepsilon\) across alternative hypotheses. See Appendix~\ref{simulation:dgp} for full details.

\begin{figure}[h!] 
    \centering
    \begin{subfigure}{0.49\textwidth}
        \centering
        \includegraphics[width=\textwidth]{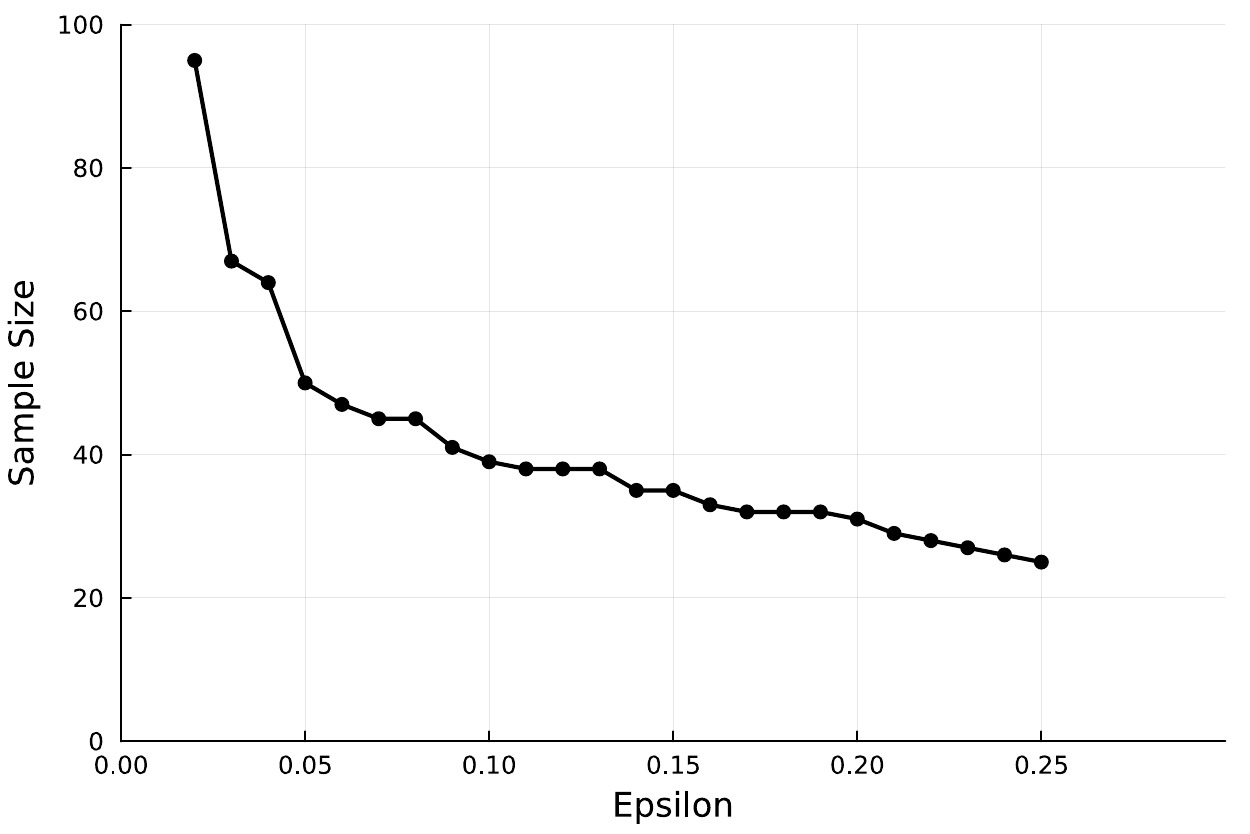}
        \caption{SUM}
        \label{fig:powerGARP}
    \end{subfigure}
    \hfill
    \begin{subfigure}{0.49\textwidth}
        \centering
        \includegraphics[width=\textwidth]{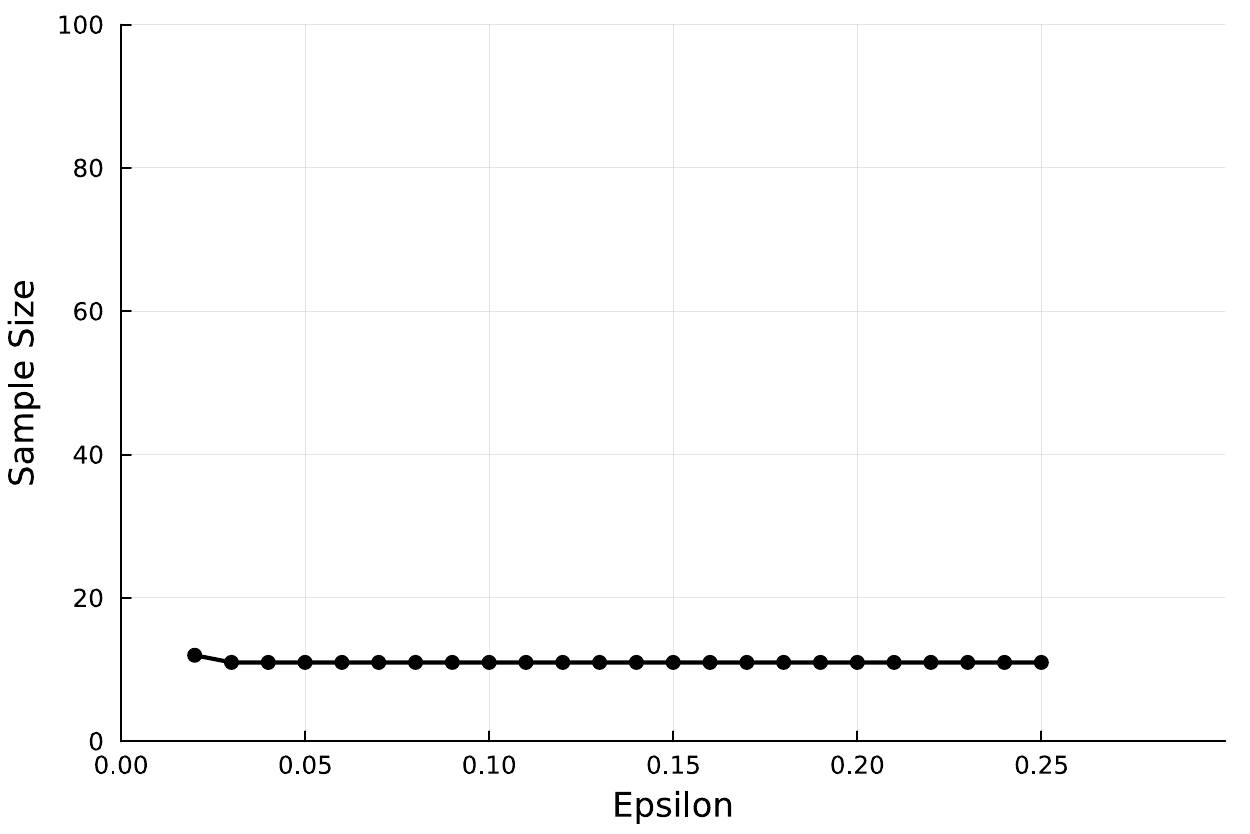}
        \caption{H}
        \label{fig:powerHARP}
    \end{subfigure}

    \vspace{1em}

    \begin{subfigure}{0.49\textwidth}
        \centering
        \includegraphics[width=\textwidth]{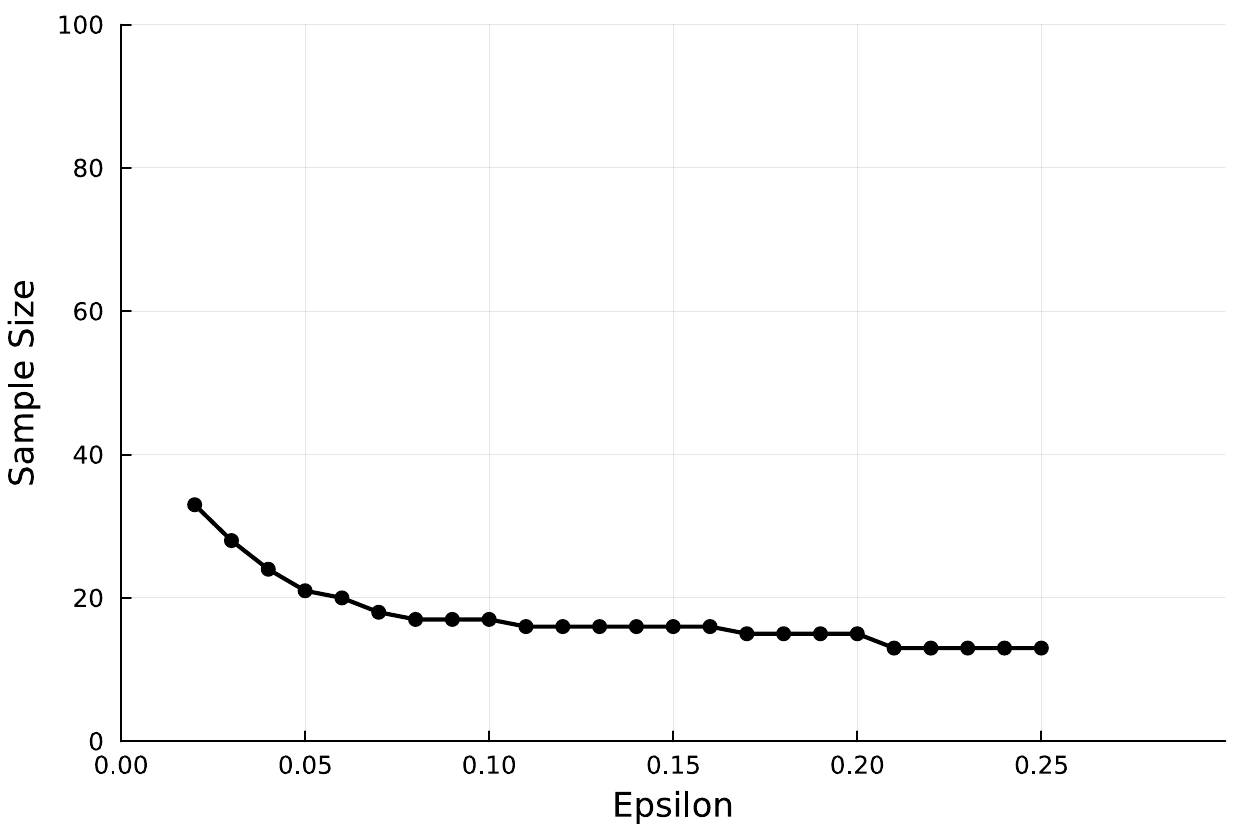}
        \caption{WS}
        \label{fig:powerWS}
    \end{subfigure}
    \hfill
    \begin{subfigure}{0.49\textwidth}
        \centering
        \includegraphics[width=\textwidth]{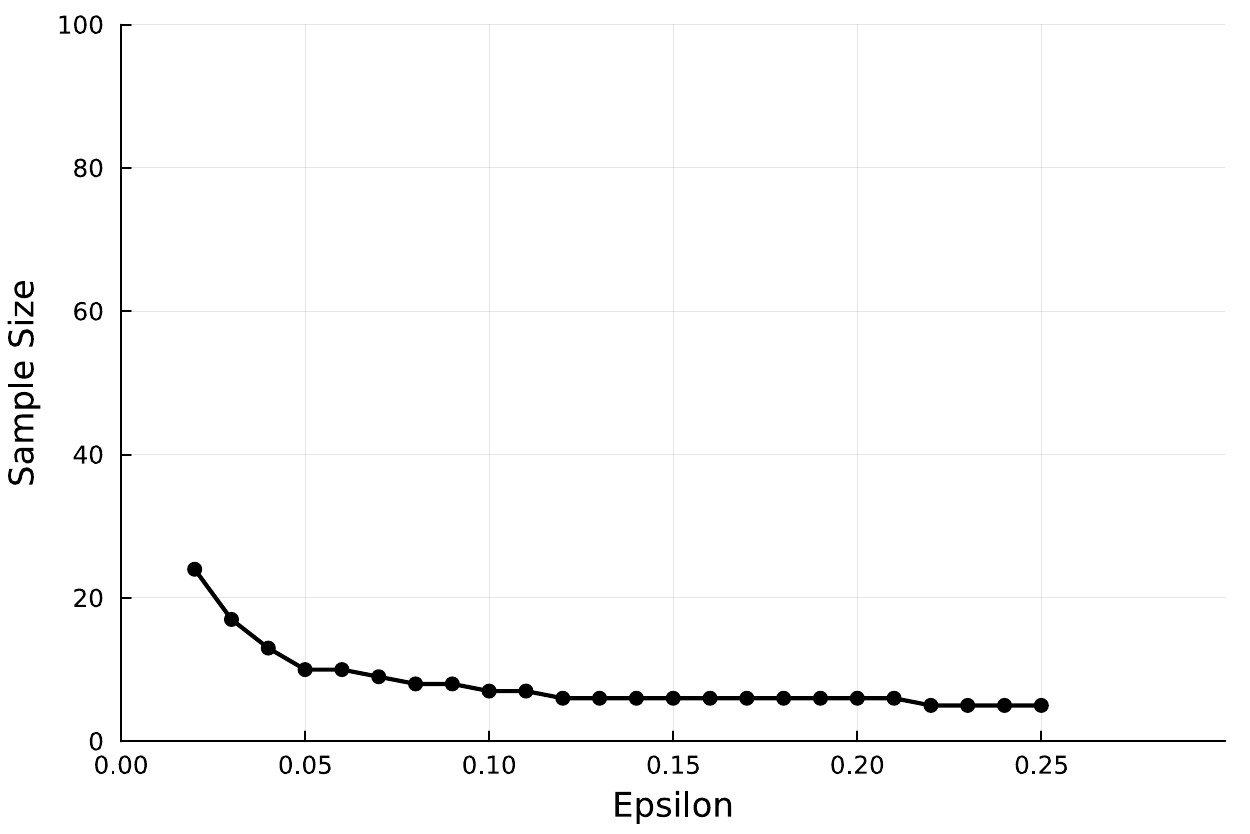}
        \caption{EU}
        \label{fig:powerEU}
    \end{subfigure}

    \caption{Power curves for SUM, H, WS, and EU.}
    \label{fig:power_all}
\end{figure}

We simulate 500 perturbed demands from Cobb-Douglas preferences drawn uniformly from the unit interval and scaled such as to sum up to one. Prices and income are drawn uniformly from $(0.01,10.00)$ and $(1.00,6.00)$, respectively. This level of price variation exceeds what is typically observed in empirical datasets. As a result, the RP tests are given favorable conditions for detecting deviations from rationality. For each value of $\epsilon \in [0.02, 0.03, \dots, 0.25]$, we compute the smallest sample size required to achieve a rejection probability of at least $90\%$. The resulting power curves are reported in Figure~\ref{fig:power_all}.\footnote{We obtain qualitatively similar power curves with an alternative data generating process featuring a different pattern of deviations from Slutsky symmetry, indicating that the relative ranking of tests is robust across alternative hypotheses.}

The figure shows that for small deviations from rationality, SUM and WS require large sample sizes to achieve the target rejection rate. In contrast, H and EU only require modest sample sizes to achieve the target rejection rate. For large deviations from rationality, every RP test achieves the target rejection rate with small to moderate sample sizes.

Next, we summarize the relationship between the distance from rationality and the sample size needed to achieve 90\% power using a log-log regression:
\begin{equation*}
    \log(\overline{n}^m) = \beta_0^m + \beta_1^m \log \varepsilon^m + \omega^m,
\end{equation*}

\noindent
where $\overline{n}^m$ is the smallest sample size needed to achieve $90\%$ power for model $m \in \{\text{SUM}, \text{H}, \text{WS}, \text{EU}\}$, $\epsilon^{m}$ is the distance from rationality, and $\omega^{m}$ captures residual variation due to the simulation process. The intercepts ($\beta_0$) reflect baseline differences in the sample size needed to reach a given power level when the deviation from rationality is one. The slopes ($\beta_1$) reflect the elasticity of the required sample size with respect to deviations from rationality. Notice that by Lemma \ref{thm: beigmanvohraPACleanrability}, fixing $\delta$ to 0.1, the order of $\beta_1$ should be smaller  than $2+l$. We find this empirically, and the results are presented in Table \ref{tab:combined_power}.\footnote{Standard errors are omitted because the estimates are derived from Monte Carlo simulations rather than empirical sampling. We show two further regressions in Appendix \ref{simulation:dgp} for completeness.}
\begin{table}[h!]
\centering
\caption{Log-log regression estimates from power curves}
\label{tab:combined_power}
\begin{tabular}{lcccc}
\toprule
Parameter & SUM & H & WS & EU \\
\midrule
$\beta_0$ & 2.6448 & 2.3719 & 2.0862 & 0.7681 \\
$\beta_1$ & -0.4573 & -0.0135 & -0.3363 & -0.5589 \\
\bottomrule
\end{tabular}
\end{table}

Table \ref{tab:combined_power} shows that the required sample size decreases progressively from SUM to H, then WS, and finally EU. The table also shows that SUM and EU have higher sensitivity of required sample sizes to deviations from rationality than H and WS. Since sample size is modeled on a log scale, EU has a steeper slope ($\beta_1$) in terms of relative changes even though its curve appears flatter in absolute sample size due to a lower baseline sample size ($\beta_0$) compared with WS.

\section{Functional Tests}\label{sec:app}

The previous section applied our framework to analyze the power of nonparametric revealed-preference tests. This section extends the analysis to tests based on functional restrictions, which may increase power or prove useful when nonparametric characterisations are difficult to implement.

\subsection{Functional Restrictions}

To develop functional tests, we first introduce \textit{functional restrictions} that single out subclasses of preferences.

\begin{mydef}[Functional Restriction]\label{def:functional-restriction}
Let $\mathcal{F}\subseteq \mathcal{G}$ be a class of demand functions. 
A map $\mathcal{R}:\mathcal{G}\rightarrow \mathbb{R}_{\geq 0}$ is called a 
\emph{functional restriction} for $\mathcal{F}$ if $\mathcal{R}(g)=0 \iff g \in \mathcal{F}$.

\end{mydef}
\noindent
Suppose we want to describe a hypothesis test of $\mathcal{F}$ as in the preceding section:
\[
\mathbf{H}_0: g \in \mathcal{F}
\qquad\text{vs.}\qquad 
\mathbf{H}_1: g \in \mathcal{G}\setminus \mathcal{F}(\varepsilon),
\]
where
\[
\mathcal{F}(\varepsilon)
:= \{ g \in \mathcal{G} : \inf_{g'\in\mathcal{F}}\mathsf{erf}_\mu(g,g') \leq \varepsilon \},
\]

\noindent
as in Theorem~\ref{thm: RP-PAC}. As the examples below show, the definition of functional restriction imposes no finite-sample bounds on size and power without additional restrictions.
\begin{example}[Size Bounds]
Let $\mathcal{F}\subset\mathcal{G}$ and define the functional restriction:
\[
\mathcal{R}_{\mathrm{ind}}(g)\;:=\;\mathbf{1}\{g\notin\mathcal{F}\}.
\]
Consider any (measurable) test $\mathcal{T}:\mathcal{D}^n\to\{0,1\}$ whose decision rule depends only on $\mathcal{R}_{\mathrm{ind}}$ (e.g., “reject iff $\mathcal{R}_{\mathrm{ind}}(g)=1$”). Notice that for any $\varepsilon>0$, we can pick some $f\in \mathcal{F}$ near the boundary such that there exists $g \notin \mathcal{F}$ with $d(f,g)\leq \varepsilon$. Hence,
\[\sup_{g \in \mathbf{H}_0}\mu_{g}(\mathcal{T}^{-1}(1))=1.\]
\end{example}

\begin{example}[Power Bounds]
Let $\mathcal{F} \subset \mathcal{G}$ and define the functional restriction
\[
\mathcal{R}_{\mathrm{lin}}(g)=\gamma\, d(g,\mathcal{F}), \quad \gamma>0.
\]
Observe that no power bounds can be computed without knowledge of $\gamma$. 
To see this, suppose the analyst observes a dataset $D_n$, picks a 
rationalization $x\in\mathcal{G}$, and computes $\mathcal{R}(x)=k$. 
By definition, $d(x,\mathcal{F})=k/\gamma$. If $\gamma$ is small, 
then $d(x,\mathcal{F})$ is large and $x$ appears to lie deep in the 
alternative; if $\gamma$ is large, then $d(x,\mathcal{F})$ is small 
and $x$ appears close to the null. Since the same observed value $k$ 
is consistent with either conclusion, the analyst cannot determine 
whether to reject without a bound on $\gamma$.
\end{example}
\noindent


As the above examples illustrate, functional restrictions alone may be too weak unless they satisfy additional properties that control their behavior under small perturbations of the data. Intuitively, we want two properties: \textit{(i)} if the true preference lies in $\mathcal{F}$, then any nearby preference should nearly satisfy the restriction (continuity), and \textit{(ii)} if a preference is far from satisfying the restriction, then this separation should be detectable in terms of our error metric (regularity). We now formalise these requirements.

\begin{mydef}[Uniform Continuity]\label{def:uniform}
A restriction $\mathcal{R}$ defining $\mathcal{F}$ is \emph{uniformly continuous} if there exists
a function $\gamma:(0,\infty)\to(0,\infty)$ with $\gamma(\varepsilon)\to 0$ as $\varepsilon\to 0$, such that for every $g,g'\in\mathcal{G}$ and every $\varepsilon>0$,
\[
\mathsf{erf}_{\mu}(g,g')\;<\;\varepsilon
\quad\Longrightarrow\quad
\big|\mathcal{R}(g)-\mathcal{R}(g')\big|\;<\;\gamma(\varepsilon).
\]
\end{mydef}

\begin{mydef}[Regularity]\label{def:regular}
A restriction $\mathcal{R}$ defining $\mathcal{F}$ is \emph{regular} if there exists
a computable strictly increasing function\footnote{Here “computable’’ means that the function $\lambda(\varepsilon)$ can be obtained explicitly 
(for example, in closed form or via an algorithm) from the specification of the restriction $\mathcal{R}$ 
and the underlying environment $\mathcal{P}\times\mathcal{I}$.} $\lambda:(0,\infty)\to(0,\infty)$ such that, for every
$g\in\mathcal{G}$ and every $\varepsilon>0$,
\[
\inf_{f\in\mathcal{F}}\mathsf{erf}_{\mu}(g,f)\;>\;\varepsilon
\quad\Longrightarrow\quad
\mathcal{R}(g)\;>\;\lambda(\varepsilon).
\]
\end{mydef}\

Putting everything together, a functional restriction is simply a condition $\mathcal{R}$ that defines a subclass $\mathcal{F}\subset\mathcal{G}$ and induces a natural hypothesis test. To make such tests well behaved in our setting, we impose uniform continuity and regularity. The former guarantees the stability of the restriction under small perturbations, while the latter guarantees detectability of violations with respect to our error metric.

\begin{rem}
The functions $\gamma(\varepsilon)$ and $\lambda(\varepsilon)$ depend only on the restriction $\mathcal{R}$ and the domain $\mathcal{P}$, not on the specific preference relation or the realised sample. This uniformity ensures that the properties of continuity and regularity can be used in constructing feasible tests without requiring knowledge of the true underlying preference.
\end{rem}

\subsection{Testing Algorithm}

The preceding subsection established the role of functional restrictions in defining well–behaved subclasses 
$\mathcal{F}\subset \mathcal{G}$ and showed how uniform continuity and regularity provide the discipline needed 
for asymptotic analysis. We now turn to the construction of an explicit testing algorithm. The idea is to exploit the functional restriction $\mathcal{R}$ to obtain a computable statistic 
that separates the null $\mathbf{H}_0:g\in\mathcal{F}$ from the alternative 
$\mathbf{H}_1:g\in \mathcal{G}\setminus \mathcal{F}(\varepsilon)$. 

The algorithm below makes concrete the approach developed in Section~\ref{ss: finite-sample}. Rather than treating revealed preference tests, it delivers a constructive procedure with explicit finite–sample power bounds. It also illustrates how functional restrictions can serve as a practical testing device in settings where algebraic characterisations of the null are either unknown or computationally prohibitive.
\begin{tcolorbox}[colback=gray!5,
                  colframe=gray!60,
                  colbacktitle=gray!30,
                  coltitle=black,
                  title=Algorithm~1: Functional--Restriction Test]
\begin{enumerate}
    \item Fix a uniformly continuous restriction $\mathcal{R}$ defining a subclass 
    $\mathcal{F}\subset\mathcal{G}$, with Lipschitz constant $L$. 
    Take parameters $\varepsilon$, $\delta>0$, and let the analyst sample a dataset of size $n$.
    \item Choose $n=n(\varepsilon/t,\delta)$, where $t$ is large enough that 
    $\lambda(\varepsilon)-\gamma(\varepsilon/t)>\gamma(\varepsilon/t)$.\footnote{Such a $t$ exists as $\gamma(0)=0$.}
    \item Pick any rationalising choice function $x$ and compute $\mathcal{R}(x)$. 
    The decision rule is
    \[
      \text{Decision} =
       \begin{cases}
         \text{Reject}, & \mathcal{R}(x)>\gamma(\varepsilon/t), \\
       \text{Accept}, & \text{otherwise}.
       \end{cases}
    \]
\end{enumerate}
\end{tcolorbox}

The algorithm operationalises functional restrictions by providing a statistic $\mathcal{R}(x)$ capable of distinguishing the null from the alternative. Step~2 ensures that the sample size is sufficiently large relative to the tolerance parameters so that the separation implied by regularity dominates the continuity margin. Step~3 then reduces the test to a simple decision rule based on whether the restriction is violated beyond the uniform–continuity threshold. 
In this way, the procedure turns the abstract properties of functional restrictions into a concrete finite–sample test. The next result uses this algorithm to construct a test based on functional restrictions. 
\begin{thm}
\label{thm:functional-test}
Following Algorithm~1, the induced test based on a uniformly continuous and regular 
restriction $\mathcal{R}$ satisfies:
\begin{enumerate}
    \item \textit{Size.} Under $\mathbf{H}_0:g\in\mathcal{F}$, the probability of rejection is at most $\delta$.
    \item \textit{Power.} Under any $\mathbf{H}_1:g\in\mathcal{G}\setminus\mathcal{F}(\varepsilon)$, 
    the probability of rejection is at least $1-\delta$.
\end{enumerate}
\end{thm}

Theorem~\ref{thm:functional-test} follows directly from the properties of uniform continuity and regularity. Under the null, Algorithm~1 ensures that any rationalising preference $\succeq_n$ lies within $\varepsilon/t$ of the true preference with probability at least $1-\delta$. Uniform continuity then implies $\mathcal{R}(x_{n})\leq\gamma(\varepsilon/t)$, so rejection occurs with probability at most $\delta$. 
Under the alternative, the true preference is separated from $\mathcal{F}$ by at least $\varepsilon$, and regularity guarantees $\mathcal{R}(g)\geq\lambda(\varepsilon)$. 
With probability at least $1-\delta$, the rationalising preference $\succeq_n$ selected by the analyst lies within $\varepsilon/t$ of the true preference, 
so $\mathcal{R}(x_{\succeq_n})\geq \lambda(\varepsilon)-\gamma(\varepsilon/t)$, which exceeds the critical value by construction. 
Hence, rejection occurs with probability at least $1-\delta$. Taken together, Algorithm~1 and Theorem~\ref{thm:functional-test} show how functional restrictions 
yield implementable frequentist tests with explicit finite–sample bounds.

\begin{rem}
    It is possible to show that our approach depends only on the learnability of the induced demand system, not on the particular type of choice sets. Once learnability is assured, the same continuity arguments and finite–sample bounds apply, so any functional restriction $\mathcal{R}$ usable with budget sets can be applied equally well to data generated from other admissible families of choice sets.
\end{rem}

\subsection{Inference for Smooth Estimands}

This section outlines how to construct confidence intervals for smooth estimands of demand. To fix ideas, let $\mathcal{R}$ denote a functional restriction that maps a demand $x_\succeq$ into a real number summarising the change in a cost–of–living index induced by the preference $\succeq$. The analyst’s goal is to use the observed data to estimate the true welfare value $\Delta W^*=\mathcal{R}(g)$, where $g$ is the demand generated by the true underlying preference. The following result shows that our methodology allows one to obtain confidence intervals on welfare estimates.

\begin{thm}\label{thm:welfare-inference}
Let $\mathcal{R}:\mathcal{G}\to\mathbb{R}$ be a uniformly continuous restriction 
with modulus $\gamma$, and let the true demand be $g\in\mathcal{G}$ with welfare 
value $\Delta W^*=\mathcal{R}(g)$. Fix $\varepsilon>0$ and $\delta\in(0,1)$. 
By sampling $n \geq n(\varepsilon,\delta)$ observations, the analyst can select 
any rationalizing demand $x$, define $\Delta W=\mathcal{R}(x)$ and construct the interval
\[
\Big[\Delta W-\gamma(\varepsilon),\;\Delta W+\gamma(\varepsilon)\Big],
\]
which contains the true value $\Delta W^*$ with probability at least $1-\delta$.
\end{thm}

Theorem~\ref{thm:welfare-inference} shows that when $\mathcal{R}$ is uniformly continuous, $x$ is ensured to lie within $\varepsilon$ of $g$, such that the values of the functional differ by at most $\varepsilon$. With probability at least $1-\delta$, the true value of the estimand lies within $\gamma(\varepsilon)$ of the computed $\Delta W$. The attainable precision depends only on the modulus of continuity of $\mathcal{R}$ and on the PAC sample complexity $n(\cdot,\cdot)$ of the class $\mathcal{G}$.
\begin{rem}
If the analyst wishes to achieve a target welfare precision $\eta>0$, 
it suffices to choose any $\varepsilon>0$ satisfying 
$\gamma(\varepsilon)\leq\eta$ and sample $n\geq n(\varepsilon,\delta)$ 
observations. Such an $\varepsilon$ exists because 
$\gamma(\varepsilon)\to 0$ as $\varepsilon\to 0$. In particular, when 
$\gamma$ is strictly increasing---as it is for all the restrictions 
constructed in Section~\ref{sec:extensions}---the analyst can simply 
set $\varepsilon=\gamma^{-1}(\eta)$ and sample 
$n\geq n(\gamma^{-1}(\eta),\delta)$.
\end{rem}

\begin{example}
    As an illustration, define the indirect utility function $V_{\succeq}(p,I) = \max u_{\succeq}(y)$ subject to $y \in B(p,I)$. For a price change $p_1$ to $p_2$, the \emph{equivalent variation} (EV) is the income adjustment after the price change that restores utility to its initial level, $V_\succeq(p_1,I) \;=\; V_\succeq\big(p_2,\, I - \mathsf{EV}_{p_1,p_2}(x_\succeq)\big)$. By \citet{hausman1981exact}, the EV is the solution to a differential equation in Marshallian demand. If $x_\succeq$ is twice continuously differentiable, then the EV functional is continuous in $x_\succeq$. Since the space $\mathcal{G}$ is compact, EV is in fact uniformly continuous, so Theorem~\ref{thm:welfare-inference} applies directly.
\end{example}

\subsection{Functional Characterization and Extensions} \label{sec:extensions}

Recall  from Section~\ref{sec:setup} that \(\mathcal{P}\times\mathcal{I}\) is compact and \(\mathcal{G}\) denotes the class of income-Lipschitz demands. 
In this section we restrict attention to the subclass \(\mathcal{G}^{R}\subseteq\mathcal{G}\) of (rational) models whose Marshallian demand \(x:\mathcal{P}\times\mathcal{I}\to Y\) is twice continuously differentiable in \((p,I)\). This additional regularity is used only to justify the derivative-based restriction operators \(\mathcal{R}\). The finite-sample size and power bounds from Sections~\ref{sec:results} are otherwise unchanged.

Throughout this section, we write
$\mathsf{D}_p x(p,I)\in\mathbb{R}^{\ell\times\ell}$ for the Jacobian with respect to prices,
with entries $(\mathsf{D}_p x)_{ij}(p,I)=\partial x_i(p,I)/\partial p_j$,
and we write $\partial_I x(p,I)\in\mathbb{R}^{\ell}$ and $\partial_I^2 x(p,I)\in\mathbb{R}^{\ell}$
for the first and second derivatives with respect to income.\footnote{Whenever the dependence on $(p,I)$ is clear, we omit the arguments.} We also use the Slutsky matrix $S(x)(p,I):=\mathsf{D}_p x(p,I)+\big(\partial_I x(p,I)\big)x(p,I)^\top$, with entries $S_{ij}(x)(p,I)=(\mathsf{D}_px)_{ij}(p,I)+x_j(p,I)\cdot\partial_I x_i(p,I)$.

\subsubsection{Homotheticity}
\label{ss:homotheticity}

Homotheticity can be tested via the linearity of Engel curves. Under our assumptions, homothetic preferences \(\succeq\) are equivalent to linearity of Marshallian demand in income at every fixed price. Equivalently, the second derivative of demand with respect to income vanishes. This yields a simple derivative-based restriction.

\begin{mydef}
    \label{def:homotheticity}
    A preference \(\succeq\) on \(Y\) is \emph{homothetic} if for all \(x,y\in Y\) and all \(t>0\), $x \succeq y$  if and only if $t x \succeq t y$.
\end{mydef}

\begin{proposition}
\label{prop:homothetic}
Let \(x_\succeq:\mathcal{P}\times\mathcal{I}\to Y\) be a \(C^2\) Marshallian demand on a compact \(\mathcal{P}\times\mathcal{I}\) arising from the rational preference \(\succeq\). Then the following are equivalent:
\begin{enumerate}\itemsep0.25em
\item \(\succeq\) is homothetic.
\item \(\partial_I^2 x_\succeq(p,I)\equiv 0\) on \(\mathcal{P}\times\mathcal{I}\).
\end{enumerate}
Consequently, the restriction operator \(\mathcal{R}^{\mathrm{hom}}(x_\succeq):=\partial_I^2 x_\succeq\) vanishes under the null of homotheticity and is uniformly continuous into \(L^1_\mu\) by Proposition~\ref{prop:UC-derivatives}.\footnote{This object is defined formally in Proposition~\ref{prop:UC-derivatives} in the appendix. For completeness, $L^1_\mu(\mathcal P\times\mathcal I)$ denotes the space of $\mu$-integrable functions on $\mathcal P\times\mathcal I$, equipped with the norm $\|f\|_{L^1_\mu}:=\int_{\mathcal P\times\mathcal I}\|f(p,I)\|\,d\mu(p,I)$.}
\end{proposition}

Homotheticity admits a transparent derivative restriction: \(\mathcal{R}^{\mathrm{hom}}(x_\succeq):=\partial_I^2 x_\succeq\) must vanish under the null. Since \(\mathcal{R}^{\mathrm{hom}}\) is uniformly continuous (Proposition~\ref{prop:UC-derivatives}), it fits directly into our testing recipe: estimate \(\mathcal{R}^{\mathrm{hom}}\) from data on \((p,I,x)\), form the empirical score, and apply Algorithm~1 to obtain finite-sample size control and power against non-homothetic alternatives.


\subsubsection{Weak Separability}

We now turn to weakly separable preferences across a given partition of goods. The revealed preference tests for this class are NP–hard, which motivates our functional approach. For simplicity, we define weak separability directly via a utility representation, as axiomatic characterisations in terms of preferences are well known (\cite{debreu60-2, Koopmans1972}).

\begin{mydef}[Weak separability]
Fix a partition \(G_1,G_2\) of the commodity index set \(\{1,\ldots,\ell\}\).
A rational preference \(\succeq\) on \(Y\) is \emph{weakly separable across \((G_1,G_2)\)} if there exist subutility indices \(v_1:\mathbb{R}^{|G_1|}_{+}\to\mathbb{R}\), \(v_2:\mathbb{R}^{|G_2|}_{+}\to\mathbb{R}\) and an aggregator \(w:\mathbb{R}^2\to\mathbb{R}\) such that $u_\succeq(x)\ =\ w\big(v_1(x_{G_1}),\,v_2(x_{G_2})\big)$ represents \(\succeq\). 
\end{mydef}

For a \(C^2\) demand \(x:\mathcal{P}\times\mathcal{I}\to Y\), the Slutsky matrix is
$\mathcal{S}(x)\ :=\ \mathsf{D}_p x\ +\ x\,(\partial_I x)^\top$, with entries \(\mathcal{S}_{ij}(x)=\partial x_i/\partial p_j + x_j\,\partial x_i/\partial I\).
Before stating the main result, we introduce some notation.  For $(p,I)$, write the between–group Slutsky block
\[
\mathcal{S}^{12}(x_{\succeq})(p,I)
:= \big(\mathcal{S}_{ij}(x_{\succeq})(p,I)\big)_{i\in G_1,\,j\in G_2}
\in \mathbb{R}^{|G_1|\times|G_2|},
\]
and the income–effect vectors
\[
a(p,I):=\big(\partial_I x_{\succeq,i}(p,I)\big)_{i\in G_1},
\qquad
b(p,I):=\big(\partial_I x_{\succeq,j}(p,I)\big)_{j\in G_2}.
\]
We also impose a two–sided income–effect nondegeneracy condition:  
for every $(p,I)$ there exist $i\in G_1$ and $j\in G_2$ with
$\partial_I x_{\succeq,i}(p,I)\neq 0$ and $\partial_I x_{\succeq,j}(p,I)\neq 0$.
\begin{proposition}
\label{prop:weak-sep-matrix}
Fix a partition $G_1,G_2\subset\{1,\ldots,\ell\}$ and let
$x_{\succeq}\in C^{2}(\mathcal{P}\times\mathcal{I};Y)$ be the Marshallian demand
generated by a rational preference $\succeq$ on $Y$. The preference $\succeq$ is weakly separable across $(G_1,G_2)$ if an only if
\begin{equation}
\label{eq:arrayWsep-matrix}
\mathcal{R}^{\mathrm{sep}} := \mathcal{S}_{ij}(x_{\succeq})\,\partial_I x_{\succeq,i'}\,\partial_I x_{\succeq,j'}
-\mathcal{S}_{i'j'}(x_{\succeq})\,\partial_I x_{\succeq,i}\,\partial_I x_{\succeq,j}=0
\end{equation}

\noindent
for all $i,i'\in G_1$ and all $j,j'\in G_2$.
\end{proposition}

The proposition shows that, under the nondegeneracy condition on income effects, weak separability across $(G_1, G_2)$ is \emph{equivalent} to the vanishing of the functional restriction $\mathcal{R}^{\mathrm{sep}}$. In other words, the structural property of weak separability is captured exactly by the condition that the between–group Slutsky block has rank one, with income effects determining the factorisation.

To see this more concretely, consider the case of three goods with $G_1=\{1\}$ and $G_2=\{2,3\}$. In this case, the between–group Slutsky block reduces to the row
\[
\mathcal{S}^{12}(x_{\succeq})(p,I) = \big[\,\mathcal{S}_{12}(x_{\succeq})(p,I)\ \ \mathcal{S}_{13}(x_{\succeq})(p,I)\,\big].
\]
The restriction operator becomes the scalar condition
\[
\mathcal{R}^{\mathrm{sep}}(x_{\succeq})(p,I)
= \mathcal{S}_{12}(x_{\succeq})(p,I)\,\partial_I x_{\succeq,3}(p,I)
- \mathcal{S}_{13}(x_{\succeq})(p,I)\,\partial_I x_{\succeq,2}(p,I).
\]
Hence, weak separability across $(\{1\},\{2,3\})$ holds if and only if
$\mathcal{R}^{\mathrm{sep}}(x_{\succeq})(p,I)=0$ for all $(p,I)$. Where both income effects
$\partial_I x_{\succeq,2}$ and $\partial_I x_{\succeq,3}$ are nonzero, this condition is
equivalent to
\[
\frac{\mathcal{S}_{12}(x_{\succeq})(p,I)}{\mathcal{S}_{13}(x_{\succeq})(p,I)}
= \frac{\partial_I x_{\succeq,2}(p,I)}{\partial_I x_{\succeq,3}(p,I)}.
\]
If one of the two income effects vanishes, the restriction reduces correspondingly to
$\mathcal{S}_{12}=0$ or $\mathcal{S}_{13}=0$. This simplified case makes transparent the role
of $\mathcal{R}^{\mathrm{sep}}$: it records the precise alignment between substitution effects
and income effects across groups that characterises weak separability.

\subsubsection{Complementarity and Substitutability}

The notions of complementarity and substitutability have been formalized in several ways in the literature.\footnote{Early definitions based on the signs of second derivatives of utility functions (see \cite{auspitz2015untersuchungen}, \cite{fisher1925mathematical}, \cite{edgeworth1897teoria}, \cite{pareto1909manuel}) are not invariant to monotone transformations. For surveys, see \cite{samuelson1974complementarity} and \cite{newman1987substitutes}.} 
Two demand–based definitions are most prominent. First, goods $i$ and $j$ are said to be \emph{gross complements} (\textit{substitutes}) whenever
\[
\frac{\partial x_i(p,I)}{\partial p_j} \;<\; 0 
\quad \Big(\,>\,0\;\Big).
\]
Because this derivative includes income effects, the notion is not symmetric. Second, following \citet{hicksallen1934}, goods $i$ and $j$ are \emph{net complements} (\textit{substitutes}) if the corresponding Hicksian demand satisfies
\[
\frac{\partial h_i(p,u)}{\partial p_j} \;<\; 0 
\quad \Big(\,>\,0\;\Big).
\]
By the Slutsky decomposition, this is equivalent to the sign of the Slutsky term
\[
\mathcal{S}_{ij}(x)(p,I)
= \frac{\partial x_i(p,I)}{\partial p_j} 
  + x_j(p,I)\,\frac{\partial x_i(p,I)}{\partial I}.
\]
Unlike the gross notion, this condition is symmetric across goods.\footnote{For a recent approach that restores intuitive appeal to these definitions, see \cite{weinstein2022direct}.} 

\begin{proposition}
\label{prop:complements}
Let $x_{\succeq}:\mathcal{P}\times\mathcal{I}\to Y$ be the Marshallian demand function 
generated by a rational preference $\succeq$ on $Y$. 
Fix distinct goods $i,j\in\{1,\ldots,\ell\}$.  
A functional restriction for \emph{gross complementarity} is
\[
\mathcal{R}^{\mathrm{grossC}}_{ij}(x_{\succeq})(p,I) 
:= \max\!\left\{0,\;\frac{\partial x_{\succeq,i}(p,I)}{\partial p_j}\right\},
\]
so that $i$ and $j$ are gross complements if and only if 
$\mathcal{R}^{\mathrm{grossC}}_{ij}(x_{\succeq})\equiv 0$. Analogous forms obtain for gross substitutability and for net complementarity/substitutability, 
replacing $\partial x_{\succeq,i}/\partial p_j$ with $-\partial x_{\succeq,i}/\partial p_j$ 
or with the Slutsky term $\mathcal{S}_{ij}(x_{\succeq})$, respectively. 
\end{proposition}

Since these functional restrictions are constructed from derivatives of uniformly continuous functions, they inherit uniform continuity away from zero. Hence, as in the case of weak separability, they provide well–behaved finite–sample tests of strict complementarity and strict substitutability within our framework.

\subsubsection{Choice under Risk}

In this section, we adapt the functional form approach of \citet[Section~III]{kubler2014asset}, 
who provide a characterization of expected utility preferences from contingent-claim demand.\footnote{See Theorem~3 in \citet[p.~3475]{kubler2014asset}.} 
Our objective is to extend this analysis to the broader \emph{betweenness class} of preferences 
introduced by \citet{dekel1986axiomatic} and \citet{chew1989axiomatic}. 
Betweenness preferences generalise expected utility by requiring indifference curves to be straight lines, though not necessarily parallel, and admit an implicit representation of utility.  

Formally, let there be $S$ states of nature, indexed by $s\in\{1,\ldots,S\}$. 
The decision maker has preferences over state–contingent consumption vectors 
$\mathbf{x}\in\mathbb{R}_{>0}^S$ and holds objective beliefs 
$\boldsymbol{\pi}=(\pi_1,\ldots,\pi_S)\in\Delta(S)$. 
Preferences are represented by a strictly increasing, thrice continuously differentiable, 
and strictly quasi–concave utility function $V(\mathbf{x};\boldsymbol{\pi})$. 
By \citet[p.~312]{dekel1986axiomatic}, a preference belongs to the betweenness class if and only if 
there exists a Bernoulli index $u:\mathbb{R}_{>0}\times[0,1]\to\mathbb{R}$, 
increasing in the first argument and continuous in the second, such that
\begin{equation}
    \label{eq:betweenness-utility}
    V(\mathbf{x};\boldsymbol{\pi})
    = \sum_{s\in S}\pi_s\,u(x_s,\,V(\mathbf{x};\boldsymbol{\pi})).
\end{equation}
We assume throughout that $u$ is thrice continuously differentiable in both arguments, concave in consumption, and satisfies $u_{11}<0$.\footnote{Here and below, subscripts denote partial derivatives of the utility index $u$: for example $u_1(x,v)=\partial u(x,v)/\partial x$ and $u_{11}(x,v)=\partial^2 u(x,v)/\partial x^2$.}

As in \citet{kubler2014asset}, we assume complete markets with strictly positive prices 
$\mathbf{p}\in\mathbb{R}_{>0}^S$ and income $I>0$. 
The Marshallian demand induced by a betweenness preference $\succeq$ is then defined by
\begin{equation}
    \label{eq:betweenness-demand}
    x_{\succeq}(\mathbf{p},I,\boldsymbol{\pi})
    \;\in\; \argmax_{\mathbf{x}\in\mathbb{R}_{>0}^S}
    \Big\{\,V(\mathbf{x};\boldsymbol{\pi}) :  \mathbf{p}\cdot\mathbf{x} \leq I \,\Big\}.
\end{equation}

\begin{proposition}
\label{prop:bet-ffr-necessity}
Let $S>2$ and let $x_{\succeq}(\mathbf{p},I,\boldsymbol{\pi})\in\mathbb{R}^S_{>0}$ be the Marshallian demand induced by a betweenness preference represented by
\[
V(\mathbf{x};\boldsymbol{\pi})=\sum_{t=1}^S \pi_t\,u\big(x_t,V(\mathbf{x};\boldsymbol{\pi})\big),
\]
with $u$ strictly increasing in consumption, $C^3$ in both arguments, concave in consumption, and $u_{11}<0$. Define $k_r(\mathbf{p},\boldsymbol{\pi}):=(\pi_r/\pi_1)(p_1/p_r)$ for $r\ge 2$. Then for each $s\in\{3,\ldots,S\}$ there exists a function
\[
f:\mathbb{R}^4_{>0}\to\mathbb{R}_{>0},\qquad
(x_1,x_2,k_2,k_s)\longmapsto f(x_1,x_2,k_2,k_s),
\]
strictly increasing in its last argument $k_s$ and satisfying $f(x,x,k_2,1)=x$ for all $x,k_2>0$, such that
\begin{equation}\label{eq:bet-f-form}
x_{\succeq,s}(\mathbf{p},I,\boldsymbol{\pi})
\;=\; f\big(x_{\succeq,1}(\mathbf{p},I,\boldsymbol{\pi}),\,x_{\succeq,2}(\mathbf{p},I,\boldsymbol{\pi}),\,k_2(\mathbf{p},\boldsymbol{\pi}),\,k_s(\mathbf{p},\boldsymbol{\pi})\big).
\end{equation}
\end{proposition}

\noindent For each fixed $s\geq 3$ define the map
\[
H_s(\mathbf{p},I;\boldsymbol{\pi}):=\Big(x_{\succeq,1}(\mathbf{p},I,\boldsymbol{\pi}),\,x_{\succeq,2}(\mathbf{p},I,\boldsymbol{\pi}),\, k_s(\mathsf{p},\boldsymbol{\pi},\, x_{\succeq,s}(\mathbf{p},I,\boldsymbol{\pi})) \Big)\in \mathbb{R}^4, 
\]
where $k_s=(\pi_s/\pi_1)(p_1/p_s)$ and $\boldsymbol{\pi}$ is treated as a parameter. Let $J_s(\mathbf{p},I)$ be the $4\times (S+1)$ Jacobian of $H_s$ with respect to $(\mathbf{p},I)$.

\begin{proposition}
\label{prop:Rbet}
Under the assumptions of Proposition~\ref{prop:bet-ffr-necessity}, for each $s\ge 3$ the Jacobian $J_s$ has rank at most $3$ at every $(\mathbf{p},I)$. Equivalently, all $4\times 4$ minors of $J_s$ vanish. Define
\[
\mathcal{R}^{\mathrm{bet}}_{s}(x_{\succeq})(\mathbf{p},I,\boldsymbol{\pi})
\;:=\;
\sum_{M\in\mathscr{M}_{4\times 4}(J_s(\mathbf{p},I))} \big(\det M\big)^2,
\]
where $\mathscr{M}_{4\times 4}(J_s)$ is the collection of all $4\times 4$ submatrices of $J_s$.
Then
\[
\mathcal{R}^{\mathrm{bet}}_{s}(x_{\succeq})\equiv 0
\quad\text{is a necessary condition for betweenness rationalisability.}
\]
\end{proposition}

Proposition~\ref{prop:Rbet} imposes a clear testable implication. 
Whenever we perturb prices and income so that, to first order, $x_1$, $x_2$, and the ratio 
$k_s=(\pi_s/\pi_1)(p_1/p_s)$ remain unchanged, then the demand for state $s$ must also remain 
unchanged. Equivalently, the only channels through which $x_s$ is allowed to move are movements 
in $x_1$, $x_2$, or $k_s$. There is no independent direction in $(\mathbf{p},I)$ that can change 
$x_s$ while holding these other quantities fixed. This is the differential analogue of the 
functional dependence in Proposition~\ref{prop:bet-ffr-necessity}: under betweenness, once 
$(x_1,x_2,k_s)$ are fixed, $x_s$ is pinned down.

The key difference with expected utility is that, under betweenness, $u_1(\cdot,V)$ depends jointly on consumption and the \emph{endogenous} 
fixed point $V$. In expected utility, $u_1(\cdot)$ is univariate, which lets the FOCs collapse 
to a representation $x_s=f(x_1,k_s)$ that characterises the model. In contrast, for 
betweenness the FOCs imply that $x_s$ is a function of $(x_1,x_2,k_2,k_s)$, but this 
\emph{does not} guarantee that a given $f$ arises from some $u(\cdot,\cdot)$ consistent with 
the fixed–point structure. Thus, necessity holds---any betweenness rationalisation must 
satisfy the rank condition--- but sufficiency need not hold in general, as not every solution to the rank 
condition corresponds to some betweenness utility.  

\begin{corollary}
\label{coro:betweeness}
Let $\mathcal{F}_{\mathrm{bet}}\subset\mathcal{G}$ denote the class of preferences satisfying 
$\mathcal{R}^{\mathrm{bet}}_{s}(x_{\succeq})\equiv 0$ for every $s\ge 3$. 
Then the class of betweenness preferences is contained in $\mathcal{F}_{\mathrm{bet}}$.
\end{corollary}

The corollary makes precise the one–sided nature of the restriction: any demand violating 
$\mathcal{R}^{\mathrm{bet}}_{s}\equiv 0$ cannot come from betweenness, but some 
non–betweenness demands may also satisfy it.

Although $\mathcal{R}^{\mathrm{bet}}$ is not a full characterisation, it yields a valid 
non–rejection test. We test
\[
\mathbf{H}_0:\ x_{\succeq}\in\mathcal{F}_{\mathrm{bet}}
\qquad\text{vs}\qquad
\mathbf{H}_1:\ x_{\succeq}\in\mathcal{G}\setminus \mathcal{F}_{\mathrm{bet}}(\varepsilon),
\]
where $\mathcal{F}_{\mathrm{bet}}(\varepsilon)$ is defined as in Section~\ref{sec:results}. For each $s\ge 3$ and each evaluation point $(\mathbf{p},I)$, compute the Jacobian 
$J_s(\mathbf{p},I)$ of 
\[
H_s(\mathbf{p},I;\boldsymbol{\pi})
=\big(x_{\succeq,1},\,x_{\succeq,2},\,k_s,\,x_{\succeq,s}\big),
\qquad k_s=(\pi_s/\pi_1)(p_1/p_s),
\]
with respect to $(\mathbf{p},I)$. Form
\[
\mathcal{R}^{\mathrm{bet}}_{s}(x_{\succeq})(\mathbf{p},I,\boldsymbol{\pi})
=\sum_{M\in\mathscr{M}_{4\times 4}(J_s(\mathbf{p},I))}(\det M)^2,
\]
and aggregate across $s$ and $(\mathbf{p},I)$ via the $L^1_\mu$–integral of 
Section~\ref{sec:efficiency}, obtaining the sample analogue $\widehat{\mathcal{R}}^{\mathrm{bet}}$. 
Under $\mathbf{H}_0$, the restriction equals zero. By Proposition~\ref{prop:UC-derivatives} and 
compactness, the map $x\mapsto \mathcal{R}^{\mathrm{bet}}$ is uniformly continuous into $L^1_\mu$, 
so the finite–sample bounds established earlier apply.  

Operationally, one can approximate the minors of $J_s$ by finite differences, using small 
perturbations $(\Delta\mathbf{p},\Delta I)$ chosen so that $x_1$, $x_2$, and $k_s$ are 
(approximately) held constant to first order. The restriction requires the induced change in $x_s$ 
to be (approximately) zero. Equivalently, the estimated $4\times 4$ minors evaluated at these 
perturbations should be close to zero under $\mathbf{H}_0$.

\section{Efficiency Gains: Queries and Subclasses}
\label{sec:efficiency}

The sample complexity of both testing and estimation in our framework is inherited from the PAC--learnability of $\mathcal{G}$. This means that the attainable precision and confidence of welfare estimates are directly tied to the sample complexity of the underlying learning algorithm. In this section, we formalise this dependence and show how efficiency gains can be realised in two ways: by imposing additional structure on the class $\mathcal{G}$, and by allowing the analyst to select prices adaptively. The following proposition makes precise the link between improved sample complexity and tighter confidence intervals.

\begin{prop}
\label{prop:efficiency-sample-complexity}
Fix a sample size $n$ and confidence level $\delta\in(0,1)$. 
Let $n_1(\varepsilon,\delta)$ and $n_2(\varepsilon,\delta)$ be two valid sample 
complexity functions for $(\varepsilon,\delta)$--learning $\mathcal{G}$ such that $n_1(\varepsilon,\delta) < n_2(\varepsilon,\delta)$ for all $\varepsilon$, $\delta$. Then the minimal confidence interval width implied by $n_1$,
\[
t_1^*(\delta,n)\ :=\ \inf\{\, t>0 : n_1(\gamma(t),\delta)\leq n \,\},
\]
is weakly smaller than that implied by $n_2$, 
\[
t_2^*(\delta,n)\ :=\ \inf\{\, t>0 : n_2(\gamma(t),\delta)\leq n \,\}.
\]
\end{prop}

Proposition~\ref{prop:efficiency-sample-complexity} formalises the connection between sample complexity and inferential precision in our framework. The sample complexity $n(\varepsilon,\delta)$ captures how demanding it is to learn the class $\mathcal{G}$: smaller values mean that fewer observations are required to reach a given accuracy and confidence. Thus, the results states that improvements in sample complexity translate directly into tighter confidence bounds at the same $(n,\delta)$. 

\begin{example}
Suppose two candidate sample complexities are given by 
\[
n_1(\varepsilon,\delta)\;=\;\frac{1}{\varepsilon^2}\cdot \frac{1}{\delta},
\qquad 
n_2(\varepsilon,\delta)\;=\;\frac{1}{\varepsilon^{k+2}}\cdot \frac{1}{\delta},
\]
with $\gamma(t)=t$. 
For each $i=1,2$, the minimal interval width is $t_i^*(\delta,n) = \inf\{\,t>0:\ n_i(\gamma(t),\delta)\leq n\,\}$. Solving these inequalities yields
\[
t_1^*(\delta,n)\;=\;(\delta n)^{-1/2},
\qquad
t_2^*(\delta,n)\;=\;(\delta n)^{-1/(k+2)}.
\]
Thus the first learning algorithm delivers strictly tighter confidence intervals, 
since $t_1^*(\delta,n)<t_2^*(\delta,n)$ for all $k>0$.

A similar comparison applies to the attainable confidence levels when the interval 
width $t$ is fixed. Define $1-\delta_i^*(t,n)\;=\;\sup\{\,1-\delta:\ n_i(\gamma(t),\delta)\leq n\,\}$. Then
\[
1-\delta_1^*(t,n)\;=\;1-\frac{1}{n t^2},
\qquad
1-\delta_2^*(t,n)\;=\;1-\frac{1}{n t^{k+2}}.
\]
Here too the algorithm with lower sample complexity, $n_1$, yields higher 
confidence at fixed precision. In Appendix \ref{sec:efficiency-appendix}, we formalize the efficiency gains that occur through adaptive sampling and further assumptions on $\mathcal{G}$.
\end{example}

\section{Conclusion} \label{sec:conclusion}

This paper develops a frequentist framework for conducting power analysis of revealed preference tests using finite choice data and tackles a central challenge in the literature: constructing parsimonious alternative hypotheses against which the discriminatory power of rationality tests can be assessed. Our procedure is grounded in the Probably Approximately Correct (PAC) learning framework and allows us to design tests that integrate functional and finite-data approaches. This method is broadly applicable, ranging from weak separability to choice under risk, highlighting its versatility for analyzing decision-making.

We make four main contributions. First, we show that PAC learnability results for Lipschitz demand functions provide the theoretical foundation for constructing well-defined alternative hypotheses. By exploiting the fact that any rationalizing demand function lies within a learnable neighborhood of the true demand with high probability, we can compute explicit bounds on the sample size required to achieve any desired level of statistical power against alternatives that are $\varepsilon$-separated from the null hypothesis.

Second, we extend our framework beyond traditional revealed preference characterizations to functional restrictions. This extension is particularly valuable for preference classes—such as weakly separable preferences—where exact RP tests are computationally intractable (NP-hard). We show that by tolerating a small but controlled size, one can construct polynomial-time tests with explicit finite-sample power guarantees.

Third, we demonstrate how to construct confidence intervals for smooth functionals of demand, such as equivalent variation and other welfare measures. The width of these intervals depends directly on the sample complexity of learning the underlying preference class, thereby linking the richness of the maintained hypothesis to the precision of welfare inference.

Fourth, our simulation results reveal an important insight: the apparent empirical success of GARP may partly reflect its limited ability to detect small deviations from rationality. In contrast, the frequent rejections of expected utility in experimental data may reflect its sensitivity to even negligible departures from the model. This suggests that comparing rejection rates across different preference classes requires careful attention to their differential statistical power. 


Our approach shifts the focus from testing the dataset itself to testing the decision maker. Unlike RP tests that always reject a dataset if it cannot be rationalized by any preference from the class of interest, our tests are one-sided: they never reject a rationalizable dataset. Still, they may fail to reject a non-rationalizable one. However, the probability of failing to reject a non-rationalizable dataset decreases as more data are sampled. Our simulation exercises show that our tests perform well in finite samples and correctly reject demands that do not belong to specific classes of preferences.

\bibliography{Raghav}


\clearpage
\begin{appendices}

\begin{center}
    \section*{\Large{Appendices (Online-Only)}}
\end{center}
\section{Efficiency Gains: Queries and Subclasses}
\label{sec:efficiency-appendix}

\subsection{Subclasses}

A natural way to obtain sharper precision bounds (and higher power for RP characterisations) 
is to impose additional structure on the preference family $\mathcal{G}$. Restricting attention to subclasses can dramatically reduce sample complexity, and hence 
tighten the confidence intervals described in Proposition~\ref{prop:efficiency-sample-complexity}. 
\cite{balcan2014learning} provide sharp results for several economically relevant subclasses, including linear preferences, separable piecewise linear concave (SPLC) 
preferences with $l$ segments, and CES preferences with a known parameter $\rho$.

To illustrate how subclass restrictions operate in practice, we recall three canonical families of preferences studied in \cite{balcan2014learning}. Each imposes a different structure on the utility function, which in turn reduces the statistical complexity of learning relative to the unrestricted class $\mathcal{G}$. 
The cases of linear, separable piecewise linear concave (SPLC), and constant elasticity of substitution (CES) preferences are of particular interest, both because of their economic relevance and because their structural features lead to sharply different sample complexities.

\begin{mydef}[Linear Preference]
A utility function $U$ is called \emph{linear} if utility is linear in each good. 
Formally, for some $a\in\mathbb{R}^d_+$,
\[
U(x)=U_a(x)=\sum_{j=1}^da_jx_j,
\]
with the normalisation $\sum_{j}a_j=1$ taken without loss of generality. 
\end{mydef}

\begin{mydef}[Separable Piecewise Linear Concave (SPLC)]
A utility function $U$ is called \emph{SPLC} if 
\[
U(x)=\sum_{j=1}^dU_j(x_j),
\]
where each $U_j:\mathbb{R}_+\to\mathbb{R}_+$ is non-decreasing, concave, and piecewise linear. 
The number of segments of $U_j$ is denoted $|U_j|$, and the $k$th segment is written $(j,k)$. 
If segment $(j,k)$ has domain $[a,b]\subseteq\mathbb{R}_+$ and slope $c$, we set $a_{jk}=c$ 
and $\ell_{jk}=b-a$, with $\ell_{j|U_j|}=\infty$. Concavity implies 
$a_{j(k-1)}>a_{jk}$ for all $k\ge 2$. 
An SPLC function with $|U_j|\le \kappa$ for all $j$ can thus be represented by two 
matrices $A,L\in\mathbb{R}_+^{d\times\kappa}$, and denoted $U_{A,L}$. 
\end{mydef}

\begin{mydef}[Constant Elasticity of Substitution (CES)]
A utility function $U$ is called \emph{CES} if, for some $\rho\in(-\infty,1]$ 
and $a\in\mathbb{R}^d_+$,
\[
U(x)=U_{a,\rho}(x)=\Bigl(\sum_{j=1}^d a_jx_j^\rho\Bigr)^{1/\rho},
\]
with the normalisation $\sum_{j}a_j=1$ taken without loss of generality. 
\end{mydef}

The efficiency gains from subclass restrictions can be quantified directly through the PAC--learning sample complexity of these classes. 
\cite{balcan2014learning} derive bounds for linear, SPLC, and CES preferences, showing how the additional structure substantially reduces the number of observations 
needed to achieve accuracy $\varepsilon$ and confidence $1-\delta$. Table~\ref{tab:sample-complexity} summarises their results.
\begin{table}[h!]
\centering
\caption{Sample Complexity by Preference Class\label{tab:sample-complexity}}
\resizebox{\textwidth}{!}{%
\begin{tabular}{@{}lccc@{}}
\toprule
\textbf{Preference Class} & \textbf{Linear} & \textbf{SPLC ($l$ segments)} & \textbf{CES} \\
\midrule
\textbf{Sample Complexity} 
& $O\left(\frac{k\log(1/\varepsilon) + \log(1/\delta)}{\varepsilon}\right)$ 
& $O\left(\frac{kl\log(1/\varepsilon) + \log(1/\delta)}{\varepsilon}\right)$ 
& $O\left(\frac{k\log(1/\varepsilon) + \log(1/\delta)}{\varepsilon}\right)$ \\
\bottomrule
\end{tabular}%
}
\end{table}

These bounds highlight how subclass assumptions translate into improved statistical performance within our framework. In particular, the dependence on $1/\varepsilon$ improves from quadratic in the unrestricted case to linear under these structured classes. This reduction feeds directly into the confidence bounds of Proposition~\ref{prop:efficiency-sample-complexity}: by replacing the generic sample complexity $n(\varepsilon,\delta)$ with the subclass--specific expressions from Table~\ref{tab:sample-complexity}, the analyst obtains strictly tighter confidence intervals and higher power for the same dataset. We next show how adaptive sampling can generate further efficiency gains.

\subsection{The Adaptive Model}

A further efficiency gain arises when the analyst is allowed to choose budget sets adaptively. In this model, the analyst selects prices sequentially and observes demand 
from the corresponding budget set, with each subsequent query conditioned on the data revealed in earlier rounds.

The effect of adaptivity on sample complexity is substantial. \cite{balcan2014learning} show that when queries are chosen adaptively, the number of observations needed to learn certain subclasses of preferences drops 
dramatically relative to the non-adaptive case. 
Table~\ref{tab:adaptive-sample-complexity} reports their bounds for linear, SPLC, 
and CES preferences. 

\begin{table}[h!]
\centering
\caption{Sample Complexity by Preference Class\label{tab:adaptive-sample-complexity}}
\begin{tabular}{@{}lccc@{}}
\toprule
\textbf{Preference Class} & \textbf{Linear} & \textbf{SPLC ($l$ segments)} & \textbf{CES} \\
\midrule
\textbf{Sample Complexity} 
& $O\left(qd\right)$ 
& $O\left(qkd\right)$ 
& $O\left(1\right)$ \\
\bottomrule
\end{tabular}
\end{table}

Here, $q$ denotes the bit-size of the coefficients. If utility functions are $L$--Lipschitz, one can interpret $q$ as setting the granularity of an approximation: 
coefficients take values in $[0,L]$ with increments of $L/(2^q-1)$. The table shows that adaptivity eliminates the $1/\varepsilon$ dependence altogether: learning becomes essentially finite in the CES case, linear in dimension for linear preferences, and only modestly larger for SPLC. 
This highlights how adaptive sampling can deliver dramatic efficiency gains in revealed--preference analysis.

\section{Proofs}
\label{app: Proofs}

\subsubsection*{A continuity proposition for derivative–based restrictions} 

Recall from Section~\ref{sec:setup} that \(\mathcal{P}\times\mathcal{I}\) is compact and \(\mathcal{G}\) denotes the class of income-Lipschitz demands. 
In this section we restrict attention to the subclass \(\mathcal{G}^{R}\subseteq\mathcal{G}\) of models whose Marshallian demand \(x:\mathcal{P}\times\mathcal{I}\to\mathbb{R}^\ell\) is twice continuously differentiable in \((p,I)\).
This additional regularity is used only to justify the derivative-based restriction operators \(\mathcal{R}\). The finite-sample size and power bounds from Sections~\ref{sec:results}–\ref{sec:efficiency} are otherwise unchanged.

\begin{proposition}
\label{prop:UC-derivatives}
Let \(\mu\in\Delta(\mathcal{P}\times\mathcal{I})\) and \(\mathcal{G}\subset C^2(\mathcal{P}\times\mathcal{I};\mathbb{R}^\ell)\).
Define \(\mathcal{R}_p(x):=\mathsf{D}_p x\), \(\mathcal{R}_I(x):=\partial_I x\), and \(\mathcal{R}_{II}(x):=\partial_I^2 x\), viewed as elements of \(L^1_\mu(\mathcal{P}\times\mathcal{I})\) endowed with the norm
\[
\|f\|_{L^1_\mu}:=\int_{\mathcal{P}\times\mathcal{I}}\|f(p,I)\|\,\mathrm{d}\mu(p,I).
\]
Then \(\mathcal{R}_p,\mathcal{R}_I:(\mathcal{G},\|\cdot\|_{C^{1,\infty}(\mathcal{P}\times\mathcal{I})})\to L^1_\mu(\mathcal{P}\times\mathcal{I})\) and
\(\mathcal{R}_{II}:(\mathcal{G},\|\cdot\|_{C^{2,\infty}(\mathcal{P}\times\mathcal{I})})\to L^1_\mu(\mathcal{P}\times\mathcal{I})\) are bounded linear maps with operator norm at most \(1\).
In particular, for all \(x,y\in\mathcal{G}\),\footnote{For \(x:\mathcal{P}\times\mathcal{I}\to\mathbb{R}^\ell\), we use the Euclidean norm \(\|\cdot\|\) on \(\mathbb{R}^\ell\).
For matrices \(A\in\mathbb{R}^{\ell\times\ell}\), we use the induced operator norm \(\|A\|:=\sup_{\|v\|=1}\|Av\|\).
The \(C^1\) sup–norm is
\[
\|x\|_{C^{1,\infty}}:=\max\Big\{\sup_{(p,I)}\|x(p,I)\|,\ \sup_{(p,I)}\|\mathsf{D}_p x(p,I)\|,\ \sup_{(p,I)}\|\partial_I x(p,I)\|\Big\},
\]
and \(\|x\|_{C^{2,\infty}}\) is defined analogously by additionally taking the sup–norm of \(\partial_I^2 x\) (and, if needed, of the remaining second and mixed partial derivatives).}
\begin{align*}
\|\mathcal{R}_p(x)-\mathcal{R}_p(y)\|_{L^1_\mu}\le \|x-y\|_{C^{1,\infty}},\quad &
\|\mathcal{R}_I(x)-\mathcal{R}_I(y)\|_{L^1_\mu}\le \|x-y\|_{C^{1,\infty}},\\
\|\mathcal{R}_{II}(x)-\mathcal{R}_{II}(y)\|_{L^1_\mu}&\le \|x-y\|_{C^{2,\infty}}.
\end{align*}
\end{proposition}

\begin{proof}
Recall that $\Omega:=\mathcal{P}\times\mathcal{I}$ is compact. We equip $\mathbb{R}^{\ell}$ with the Euclidean norm $\|\cdot\|$, matrices in $\mathbb{R}^{\ell\times\ell}$ with the induced operator norm $\|A\|:=\sup_{\|v\|=1}\|Av\|$, and higher-order derivatives with the corresponding multilinear operator norms.

For $x:\Omega\to\mathbb{R}^{\ell}$ define
\[
\|x\|_{C^{1,\infty}(\Omega)}
:= \max\Big\{\,\sup_{(p,I)\in\Omega}\|x(p,I)\|,\ \sup_{(p,I)\in\Omega}\|\mathsf{D}_p x(p,I)\|,\ \sup_{(p,I)\in\Omega}\|\partial_I x(p,I)\|\,\Big\},
\]
and
\[
\|x\|_{C^{2,\infty}(\Omega)}
:= \max\Big\{\,\|x\|_{C^{1,\infty}(\Omega)},\ \sup_{(p,I)\in\Omega}\|\partial_I^2 x(p,I)\|,\ \sup_{(p,I)\in\Omega}\|\mathsf{D}_p\partial_I x(p,I)\|,\ \sup_{(p,I)\in\Omega}\|\mathsf{D}_p^2 x(p,I)\|\,\Big\},
\]
where $\big(\mathsf{D}_p\partial_I x(p,I)\big)_{ij}=\partial^2 x_i(p,I)/(\partial p_j\,\partial I)$.

For any measurable vector or matrix valued function $f$ on $\Omega$, define
\[
\|f\|_{L^1_\mu(\Omega)} \ :=\ \int_{\Omega}\|f(p,I)\|\,\mathrm{d}\mu(p,I).
\]
Since $\mu\in\Delta(\Omega)$ is a probability measure, $\mu(\Omega)=1$, and therefore
\[
\|f\|_{L^1_\mu(\Omega)} \ \le\ \|f\|_{L^\infty(\Omega)}
:=\sup_{(p,I)\in\Omega}\|f(p,I)\|.
\]

\smallskip

\emph{(i) Boundedness of $\mathcal{R}_p$ and $\mathcal{R}_I$.}
Define the range spaces
\[
\mathcal{R}_p:\mathcal{G}\to L^1_\mu(\Omega;\mathbb{R}^{\ell\times\ell}),
\qquad
\mathcal{R}_I:\mathcal{G}\to L^1_\mu(\Omega;\mathbb{R}^{\ell}),
\]
by $\mathcal{R}_p(x):=\mathsf{D}_p x$ and $\mathcal{R}_I(x):=\partial_I x$.
Linearity is immediate from linearity of differentiation. For $x,y\in\mathcal{G}$,
\begin{align*}
\|\mathcal{R}_p(x)-\mathcal{R}_p(y)\|_{L^1_\mu(\Omega)}
&=\int_{\Omega}\|\mathsf{D}_p(x-y)(p,I)\|\,\mathrm{d}\mu(p,I)
\le \|\mathsf{D}_p(x-y)\|_{L^\infty(\Omega)}
\le \|x-y\|_{C^{1,\infty}(\Omega)},
\\
\|\mathcal{R}_I(x)-\mathcal{R}_I(y)\|_{L^1_\mu(\Omega)}
&=\int_{\Omega}\|\partial_I(x-y)(p,I)\|\,\mathrm{d}\mu(p,I)
\le \|\partial_I(x-y)\|_{L^\infty(\Omega)}
\le \|x-y\|_{C^{1,\infty}(\Omega)}.
\end{align*}
Hence $\mathcal{R}_p$ and $\mathcal{R}_I$ are bounded linear maps with operator norm at most $1$.

\smallskip

\emph{(ii) Boundedness of $\mathcal{R}_{II}$.}
Define
\[
\mathcal{R}_{II}:\mathcal{G}\to L^1_\mu(\Omega;\mathbb{R}^{\ell}),
\qquad
\mathcal{R}_{II}(x):=\partial_I^2 x.
\]
Again, linearity is immediate. For $x,y\in\mathcal{G}$,
\begin{align*}
\|\mathcal{R}_{II}(x)-\mathcal{R}_{II}(y)\|_{L^1_\mu(\Omega)}
&=\int_{\Omega}\|\partial_I^2(x-y)(p,I)\|\,\mathrm{d}\mu(p,I)
\le \|\partial_I^2(x-y)\|_{L^\infty(\Omega)}
\le \|x-y\|_{C^{2,\infty}(\Omega)}.
\end{align*}
Therefore $\mathcal{R}_{II}$ is a bounded linear map with operator norm at most $1$.
\end{proof}

\subsection{Proof of Lemma \ref{thm: beigmanvohraPACleanrability}}

Fix $L>0$ and let $\mathcal{G}$ be the class of choice functions
$x:\mathcal{P}\times\mathcal{I}\to\mathbb{R}^{\ell}$ that are $L$-Lipschitz with respect to
the Euclidean distance on $\mathcal{P}\times\mathcal{I}\subset\mathbb{R}^{\ell+1}$ and the Euclidean norm on
$\mathbb{R}^{\ell}$, that is, for all $(p,I),(p',I')\in\mathcal{P}\times\mathcal{I}$,
\[
\|x(p,I)-x(p',I')\|\le L\,\|(p,I)-(p',I')\|_2.
\]
We prove the stated sample-complexity bound by bounding the fat-shattering dimension of $\mathcal{G}$
via a packing-number argument and then invoking the fat-shattering sample-complexity theorem of
\citet{beigman2006learning}.

\smallskip

\noindent\textbf{Step 1.}
Let $\gamma>0$ and suppose that $S=\{s_1,\dots,s_n\}\subset \mathcal{P}\times\mathcal{I}$ is $\gamma$-fat shattered by $\mathcal{G}$
in the (vector-valued) sense of \citet[Definition 2]{beigman2006learning}. Then there exist witnesses
$y_1,\dots,y_n\in\mathbb{R}^{\ell}$ and two parallel affine hyperplanes $H_0,H_1\subset\mathbb{R}^{\ell}$
with $\mathrm{dist}(H_0,H_1)>\gamma$ such that for every labelling $b\in\{0,1\}^n$ there exists $x_b\in\mathcal{G}$
with $x_b(s_i)$ lying on the $b_i$-designated side of the translated hyperplane $y_i+H_{b_i}$.

Fix $i\neq j$ and choose a labelling $b$ with $b_i=0$ and $b_j=1$. By construction,
$x_b(s_i)$ lies in the half-space determined by $y_i+H_0$ and $x_b(s_j)$ lies in the opposite half-space determined by $y_j+H_1$.
Because the two hyperplanes are parallel and at distance greater than $\gamma$, this implies
\[
\|x_b(s_i)-x_b(s_j)\|>\gamma.
\]
Since $x_b$ is $L$-Lipschitz, we also have
\[
\|x_b(s_i)-x_b(s_j)\|\le L\,\|s_i-s_j\|_2.
\]
Combining the two inequalities yields $\|s_i-s_j\|_2>\gamma/L$. Since $i\neq j$ were arbitrary, the set $S$ is
$(\gamma/L)$-separated in Euclidean distance. Therefore,
\[
\mathrm{fat}_{\mathcal{G}}(\gamma)\le \mathsf{Pack}\big(\mathcal{P}\times\mathcal{I},\|\cdot\|_2,\gamma/L\big),
\]
where $\mathsf{Pack}(X,\|\cdot\|_2,r)$ denotes the $r$-packing number of $X$ under Euclidean distance.
This is the same separation argument used in the proof of \citet[Theorem 6]{beigman2006learning}, and it uses only the Lipschitz property.

\smallskip

\noindent\textbf{Step 2.}
Since $\mathcal{P}\times\mathcal{I}$ is a compact subset of $\mathbb{R}^{\ell+1}$, there exists a constant $C<\infty$
(depending only on $\mathcal{P}\times\mathcal{I}$) such that for all sufficiently small $r>0$,
\[
\mathsf{Pack}\big(\mathcal{P}\times\mathcal{I},\|\cdot\|_2,r\big)\le C\, r^{-(\ell+1)}.
\]
Applying this with $r=\gamma/L$ and combining with Step 1 gives
\[
\mathrm{fat}_{\mathcal{G}}(\gamma)\le C\left(\frac{L}{\gamma}\right)^{\ell+1}.
\]

\smallskip

\noindent\textbf{Step 3.}
By \citet[Theorem 4]{beigman2006learning}, the class $\mathcal{G}$ is PAC learnable with sample complexity
\[
m_L(\varepsilon,\delta)
=
O\left(
\frac{1}{\varepsilon^2}\left(
\ln^2\left(\frac{1}{\varepsilon}\right)\,\mathrm{fat}_{\mathcal{G}}(\varepsilon)
+\ln\left(\frac{1}{\delta}\right)
\right)
\right).
\]
Substituting the bound from Step 2 at $\gamma=\varepsilon$ yields
\[
m_L(\varepsilon,\delta)
=
O\left(
\frac{1}{\varepsilon^2}\left(
\ln^2\left(\frac{1}{\varepsilon}\right)\left(\frac{L}{\varepsilon}\right)^{\ell+1}
+\ln\left(\frac{1}{\delta}\right)
\right)
\right),
\]
which is the desired bound.

\subsection{Proof of Theorem \ref{thm: RP-PAC}}

By Lemma \ref{thm: beigmanvohraPACleanrability} (\cite{beigman2006learning}), the class $\mathcal{G}$ is PAC learnable: for each $\varepsilon,\delta>0$ there exists an algorithm $L$ with sample size $m_L(\varepsilon,\delta)$ satisfying Definition~\ref{def:PAC Learnability}. By the convention introduced after Definition~\ref{def:PAC Learnability}, $n(\varepsilon,\delta)$ denotes the \emph{minimal} sample complexity, hence
\[
n(\varepsilon,\delta)\ \le\ m_L(\varepsilon,\delta).
\]
Fix any sample size $n$ with $n\ge n(\varepsilon,\delta)$ (in particular, choosing $n=m_L(\varepsilon,\delta)$ suffices). Then, for any true model $g\in \mathcal{G}$,
\begin{equation*}
\mu^n_{g}\!\Big(\,\mathsf{erf}_\mu(x_{\succeq},g)<\varepsilon\ \text{for every }x_{\succeq}\in\mathcal{G}\ \text{that rationalises }D_n\,\Big)\ \ge\ 1-\delta.
\end{equation*}

Suppose now that the true model satisfies the alternative such that $g\in \mathcal{G}\setminus \mathcal{F}(\varepsilon)$. Then, it follows that
\begin{equation*}
    \inf_{x \in \mathcal{F}}\big\{ \mathsf{erf}_\mu(x,g)\big\}>\varepsilon.
\end{equation*}

\noindent
Consider any dataset $D_n$ generated by $g$ and suppose some $x\in \mathcal{F}$ rationalizes $D_n$. 
Since $\mathcal{F}\subseteq \mathcal{G}$, this $x$ is among the rationalisers quantified in the PAC event above. Hence, on that event, $\mathsf{erf}_\mu(x,g)<\varepsilon$.
This contradicts $\inf_{x'\in\mathcal{F}}\{ \mathsf{erf}_\mu(x',g)\}>\varepsilon$: the fact that $g$ lies at distance strictly greater than $\varepsilon$ from $\mathcal{F}$. Hence, with probability at least $1-\delta$, no element of $\mathcal{F}$ rationalizes $D_n$.\par 

Define the size-zero test $\tilde{\mathcal{T}}_{\mathcal{F}}$ that rejects exactly those datasets that admit no rationalization in $\mathcal{F}$. Then, $\tilde{\mathcal{T}}_{\mathcal{F}}(D_n)=1$ with probability at least $1-\delta$ under the alternative. By Definition \ref{def:RPcharacterization}, any RP characterization $\mathcal{T}_{\mathcal{F}}$ has maximal power among size-zero tests, and therefore $\mathcal{T}_{\mathcal{F}}(D_n)=1$ whenever $\tilde{\mathcal{T}}_{\mathcal{F}}(D_n)=1$. Thus
\begin{equation*}
    \mu^n_{g}\bigl(\mathcal{T}_{\mathcal{F}}^{-1}(1)\,\bigr)\ \ge\ 1-\delta.
\end{equation*}
This establishes that $\mathcal{T}_{\mathcal{F}}$ has power at least $1-\delta$ against all alternatives $g\in \mathcal{G}\setminus \mathcal{F}(\varepsilon)$. Moreover, by Definition \ref{def:RPcharacterization}\,\textit{(i)}, $\mathcal{T}_{\mathcal{F}}$ never rejects when $g\in \mathcal{F}$, so its size is zero.

\subsection{Proof of Theorem \ref{thm:power-from-fixed-n}}

Fix $\varepsilon>0$ and $n\in\mathbb{N}$. Let $\mathcal{T}$ be an RP characterisation of $\mathcal{F}$.
By the PAC result for $\mathcal{G}$ (Lemma~\ref{thm: beigmanvohraPACleanrability}),
for any $\delta\in(0,1)$ and any true model $g\in\mathcal{G}$, if $n\ge n(\varepsilon,\delta)$ then
\begin{equation}\label{eq:PAC}
\mu^n_{g}\Big(\,\mathsf{erf}_\mu(g,x) < \varepsilon\ \text{ for every } x\in\mathcal{G}\ \text{that rationalises } D_n\,\Big) \ \ge\ 1-\delta .
\end{equation}

\smallskip

\noindent For the fixed $\varepsilon>0$ and $n$, write
\(
S:=\{\delta\in(0,1):\ n(\varepsilon,\delta)\le n\}.
\)
By the non-triviality assumption stated in the text preceding the theorem, $S\neq\emptyset$. Define
\(
\delta(\varepsilon,n)\ :=\ \inf S.
\)
By the definition of the infimum, for every $m\in\mathbb{N}$ there exists $\delta_m\in S$ such that
\(
\delta(\varepsilon,n)\le \delta_m<\delta(\varepsilon,n)+\frac{1}{m}.
\)
Thus we may fix a sequence $(\delta_m)_{m\ge1}\subset S$ with $\delta_m\downarrow \delta(\varepsilon,n)$ and
\begin{equation}\label{eq:n-bounds}
n(\varepsilon,\delta_m)\ \le\ n \qquad \text{for all } m .
\end{equation}

\smallskip

\noindent Fix any alternative $g\in\mathcal{G}$ satisfying the weak separation from $\mathcal{F}$,
\[
\inf_{f\in\mathcal{F}} \mathsf{erf}_\mu(g,f)\ \ge\ \varepsilon .
\]
For each $m$, since \eqref{eq:n-bounds} implies $n\ge n(\varepsilon,\delta_m)$, applying \eqref{eq:PAC} at confidence level $1-\delta_m$ yields
\(
\mu^n_{g}(E_m)\ \ge\ 1-\delta_m,
\)
where
\[
E_m\ :=\ \Big\{D_n:\ \mathsf{erf}_\mu(g,x) < \varepsilon\ \text{ for every } x\in\mathcal{G}\ \text{that rationalises } D_n\Big\}.
\]

\smallskip

\noindent The remainder of the argument is identical to the corresponding step in the proof of Theorem~\ref{thm: RP-PAC}: on $E_m$ there cannot exist any $f\in\mathcal{F}$ that rationalises $D_n$, hence
\[
E_m\ \subseteq\ \{D_n:\ \text{no } f\in\mathcal{F}\text{ rationalises }D_n\}.
\]
Defining the canonical size--zero test $\tilde{\mathcal{T}}_{\mathcal{F}}(D_n):=\mathbf{1}\{\text{no } f\in\mathcal{F}\text{ rationalises }D_n\}$ and using the maximal-power property of RP characterisations (Definition~\ref{def:sizepower}\,\textit{(iii)}), we obtain for every $m$,
\[
\mu^n_{g}\!\big(\,\mathcal{T}^{-1}(1)\,\big)\ \ge\ 1-\delta_m .
\]
Letting $m\to\infty$ and using $\delta_m\downarrow \delta(\varepsilon,n)$ yields
\[
\mu^n_{g}\!\big(\,\mathcal{T}^{-1}(1)\,\big)\ \ge\ 1-\delta(\varepsilon,n).
\]
Since $g$ was an arbitrary alternative with $\inf_{f\in\mathcal{F}}\mathsf{erf}_\mu(g,f)\ge\varepsilon$,
this establishes the claimed power bound.

\subsection{Proof of Theorem \ref{thm:min-detectable-separation}}

Fix $n\in\mathbb{N}$ and $\delta\in(0,1)$ as in the non-triviality assumption stated in the text preceding the theorem, and fix an arbitrary $\gamma>0$. Let $\mathcal{T}$ be an RP characterisation of $\mathcal{F}$. Define
\[
\varepsilon(n,\delta)\;:=\;\inf\{\varepsilon>0:\ n(\varepsilon,\delta)\le n\},
\qquad
\bar\varepsilon\;:=\;\varepsilon(n,\delta)+\gamma.
\]

Write
\(
S\;:=\;\{\varepsilon>0:\ n(\varepsilon,\delta)\le n\}.
\)
By assumption, $S\neq\emptyset$. Since $\varepsilon(n,\delta)=\inf S$, by the definition of infimum there exists $\varepsilon'\in S$ such that $\varepsilon' < \bar\varepsilon$. In particular,
\(
n(\varepsilon',\delta)\le n.
\)
Moreover, because $n(\varepsilon,\delta)$ is a \emph{minimal} sample complexity, for each fixed $\delta$ the map $\varepsilon\mapsto n(\varepsilon,\delta)$ is weakly decreasing. Hence, since $\bar\varepsilon>\varepsilon'$,
\(
n(\bar\varepsilon,\delta)\ \le\ n(\varepsilon',\delta)\ \le\ n.
\)

By the PAC result for $\mathcal{G}$ (Lemma~\ref{thm: beigmanvohraPACleanrability}), for any true model $g\in\mathcal{G}$,
if $n\ge n(\bar\varepsilon,\delta)$ then
\begin{equation}\label{eq:PAC-MDE-new}
\mu^n_{g}\Big(\,
\mathsf{erf}_\mu(g,x)<\bar\varepsilon
\ \text{ for every } x\in\mathcal{G}\ \text{rationalising }D_n
\,\Big)\ \ge\ 1-\delta.
\end{equation}
Define the event
\[
E\;:=\;\Big\{D_n:\ \mathsf{erf}_\mu(g,x)<\bar\varepsilon\ \text{ for every } x\in\mathcal{G}\ \text{rationalising }D_n\Big\}.
\]
Then \eqref{eq:PAC-MDE-new} reads $\mu^n_{g}(E)\ge 1-\delta$. Now fix any alternative $g\in\mathcal{G}$ satisfying
\begin{equation}\label{eq:sep-MDE-new}
\inf_{f\in\mathcal{F}}\mathsf{erf}_\mu(g,f)\ \ge\ \bar\varepsilon.
\end{equation}

We claim that on $E$ there cannot exist any $f\in\mathcal{F}$ that rationalises $D_n$.
Indeed, suppose by contradiction that for some $D_n\in E$ there exists $f\in\mathcal{F}$ rationalising $D_n$.
Since $\mathcal{F}\subseteq\mathcal{G}$, this $f$ is among the $\mathcal{G}$-rationalisers quantified in the definition of $E$,
so $D_n\in E$ implies $\mathsf{erf}_\mu(f,g)<\bar\varepsilon$.
This contradicts \eqref{eq:sep-MDE-new}, since \eqref{eq:sep-MDE-new} implies $\mathsf{erf}_\mu(f,g)\ge \bar\varepsilon$ for every $f\in\mathcal{F}$.
Hence
\[
E\ \subseteq\ \{D_n:\ \text{no } f\in\mathcal{F}\ \text{rationalises }D_n\}.
\]
Therefore,
\[
\mu^n_{g}\Big(\{D_n:\ \text{no } f\in\mathcal{F}\ \text{rationalises }D_n\}\Big)\ \ge\ \mu^n_{g}(E)\ \ge\ 1-\delta.
\]

\noindent Define the canonical size--zero test
\[
\tilde{\mathcal{T}}_{\mathcal{F}}(D_n)\;:=\;\mathbf{1}\{\text{no } f\in\mathcal{F}\ \text{rationalises }D_n\}.
\]
By construction, $\tilde{\mathcal{T}}_{\mathcal{F}}$ never rejects under the null $x_{\succeq}\in\mathcal{F}$, so it has size zero.
Moreover, the inequality above gives
\[
\mu^n_{g}\big(\tilde{\mathcal{T}}_{\mathcal{F}}^{-1}(1)\big)\ \ge\ 1-\delta.
\]
Finally, since $\mathcal{T}$ is an RP characterisation of $\mathcal{F}$, it has maximal power among size--zero tests,
hence $\tilde{\mathcal{T}}_{\mathcal{F}}^{-1}(1)\subseteq \mathcal{T}^{-1}(1)$ and therefore
\[
\mu^n_{g}\big(\mathcal{T}^{-1}(1)\big)\ \ge\ 1-\delta.
\]
Since $g$ was an arbitrary alternative satisfying \eqref{eq:sep-MDE-new}, this completes the proof.

\subsection{Proof of Theorem \ref{thm:functional-test}}

We prove size and power separately. Throughout, let $g$ denote the true demand generated by the true preference $g$, and let $x_{\succeq_n}\in\mathcal{G}$ be any rationalising demand selected by the analyst from the dataset $D_n$. Recall that Algorithm~1 fixes $\varepsilon,\delta>0$, chooses $n=n(\varepsilon/t,\delta)$ with $t$ large enough so that $\lambda(\varepsilon)-\gamma(\varepsilon/t)>\gamma(\varepsilon/t)$, and rejects whenever $\mathcal{R}(x_{\succeq_n})>\gamma(\varepsilon/t)$.

\medskip
\noindent\textit{(i) Size.} Suppose the null holds, i.e.\ $g\in\mathcal{F}$. By the PAC result for $\mathcal{G}$ (Lemma~\ref{thm: beigmanvohraPACleanrability}), for the chosen $n=n(\varepsilon/t,\delta)$,
\begin{equation}\label{eq:size-distance}
\mu^n_{g}\Big(\,\mathsf{erf}_\mu\big(x_{\succeq_n},g\big)<\tfrac{\varepsilon}{t}\,\Big)\ \ge\ 1-\delta.
\end{equation}
Because $g\in\mathcal{F}$, we have $\mathcal{R}(g)=0$. Uniform continuity of $\mathcal{R}$ at $0$ yields
\begin{equation}\label{eq:size-cont}
\mathsf{erf}_\mu\big(x_{\succeq_n},g\big)<\tfrac{\varepsilon}{t}
\ \Longrightarrow\
\big|\mathcal{R}(x_{\succeq_n})-\mathcal{R}(g)\big|<\gamma(\varepsilon/t),
\end{equation}
hence $\mathcal{R}(x_{\succeq_n})<\gamma(\varepsilon/t)$. Thus, on the event in \eqref{eq:size-distance}, the test does \emph{not} reject. Therefore the probability of rejection under the null is at most $\delta$ for all $g\in \mathcal{F}$. Hence, the supremum of this rejection probability must be smaller than this $\delta$. 

\medskip
\noindent\textit{(ii) Power.} Suppose the alternative holds, i.e.\ $g\in\mathcal{G}\setminus\mathcal{F}(\varepsilon)$, equivalently
\begin{equation}\label{eq:alt-sep}
\inf_{x_{\succeq'}\in\mathcal{F}}\,\mathsf{erf}_\mu\big(g,x_{\succeq'}\big)\ \ge\ \varepsilon.
\end{equation}
By regularity of $\mathcal{R}$, condition~\eqref{eq:alt-sep} implies that the true demand must 
satisfy a numerical lower bound: 
\begin{equation}\label{eq:alt-R}
\mathcal{R}(g) \;\geq\; \lambda(\varepsilon).
\end{equation}
That is, whenever $g$ lies at least $\varepsilon$ away from $\mathcal{F}$ in the 
$\mathsf{erf}_\mu$ metric, the restriction functional assigns it a value no smaller than 
$\lambda(\varepsilon)$.
Again by Lemma~\ref{thm: beigmanvohraPACleanrability} with $n=n(\varepsilon/t,\delta)$,
\begin{equation}\label{eq:alt-distance}
\mu^n_{g}\!\left(\,\mathsf{erf}_\mu\big(x_{\succeq_n},g\big)<\frac{\varepsilon}{t}\,\right)\ \ge\ 1-\delta.
\end{equation}
Uniform continuity gives, on the event in \eqref{eq:alt-distance},
\begin{equation}\label{eq:alt-cont}
\big|\mathcal{R}(x_{\succeq_n})-\mathcal{R}(g)\big|<\gamma(\varepsilon/t)
\ \Longrightarrow\
\mathcal{R}(x_{\succeq_n})\ \ge\ \mathcal{R}(g)-\gamma(\varepsilon/t).
\end{equation}
Combining \eqref{eq:alt-R} and \eqref{eq:alt-cont} yields
\[
\mathcal{R}(x_{\succeq_n})\ \ge\ \lambda(\varepsilon)-\gamma(\varepsilon/t).
\]
By the choice of $t$, $\lambda(\varepsilon)-\gamma(\varepsilon/t)>\gamma(\varepsilon/t)$, so the decision rule in Algorithm~1 rejects on the event in \eqref{eq:alt-distance}. Hence the probability of rejection under the alternative is at least $1-\delta$. The two parts establish the stated size and power bounds, completing the proof.

\subsection{Proof of Theorem \ref{thm:welfare-inference}}
Fix $\varepsilon,\delta>0$ and let $n=n(\varepsilon,\delta)$.
Let $g\in\mathcal{G}$ denote the true demand and let $D_n$ be the dataset of 
size $n$ sampled from $\mu$ under $g$. Let $x_{\succeq}\in\mathcal{G}$ be any 
demand that rationalises $D_n$. By PAC learnability of $\mathcal{G}$ 
(Lemma~\ref{thm: beigmanvohraPACleanrability}),
\[
\mu^n_{g}\Big(\,
\mathsf{erf}_\mu\big(x_{\succeq},g\big)<\varepsilon
\ \text{ for every } x_{\succeq}\in\mathcal{G}
\ \text{that rationalises }D_n
\,\Big)\;\ge\;1-\delta.
\]
Since $\mathcal{R}$ is uniformly continuous, for every pair 
$(x_{\succeq},g)$ with 
$\mathsf{erf}_\mu(x_{\succeq},g)<\varepsilon$ we have
\[
\big|\mathcal{R}(x_{\succeq})-\mathcal{R}(g)\big|\;<\;\gamma(\varepsilon).
\]
Define $\Delta W:=\mathcal{R}(x_{\succeq})$ and 
$\Delta W^*:=\mathcal{R}(g)$. Then
\[
\mu^n_{g}\big(|\Delta W-\Delta W^*|<\gamma(\varepsilon)\big)
\;\ge\;1-\delta,
\]
which is the claimed finite-sample bound.

\subsection{Proof of Proposition \ref{prop:weak-sep-matrix}}

If $\succeq$ is weakly separable across $(G_1,G_2)$, then there exists a scalar $\kappa:\mathcal{P}\times\mathcal{I}\to\mathbb{R}$ such that
\begin{equation}
\label{eq:rankone-matrix}
\mathcal{S}^{12}(x_{\succeq})(p,I) = \kappa(p,I)\,a(p,I)\,b(p,I)^{\!\top}
\qquad \forall (p,I)\in\mathcal{P}\times\mathcal{I}.
\end{equation}
Equivalently, for all $i,i'\in G_1$ and $j,j'\in G_2$,
\begin{equation}
\label{eq:arrayWsep-matrix-2}
\mathcal{S}_{ij}(x_{\succeq})\,\partial_I x_{\succeq,i'}\,\partial_I x_{\succeq,j'}
-\mathcal{S}_{i'j'}(x_{\succeq})\,\partial_I x_{\succeq,i}\,\partial_I x_{\succeq,j}
=0 .
\end{equation}

\noindent
Define $\mathcal{R}^{\mathrm{sep}}(x_{\succeq})$ to collect the left–hand side of
\eqref{eq:arrayWsep-matrix}. Then weak separability implies
$\mathcal{R}^{\mathrm{sep}}(x_{\succeq})\equiv 0$. 

Conversely, suppose the nondegeneracy condition holds. 
If $\mathcal{R}^{\mathrm{sep}}(x_{\succeq})\equiv 0$, then there exists $\kappa$ such that \eqref{eq:rankone-matrix} holds. By \citet{goldman1964note}, 
this is equivalent to weak separability across $(G_1,G_2)$. Since $\mathcal{S}^{12}(x_{\succeq})$ and $a,b$ are built from $\mathsf{D}_p x_{\succeq}$ and $\partial_I x_{\succeq}$, uniform continuity of  $\mathcal{R}^{\mathrm{sep}}$ into $L^1_\mu$ follows from
Proposition~\ref{prop:UC-derivatives}. 
\subsection{Proof of Proposition \ref{prop:homothetic}}

Throughout the proof we write $x$ for the demand $x_\succeq$ generated by $\succeq$. By strong monotonicity the budget binds at every \((p,I)\), and by strict convexity the solution is unique and interior. Fix \(p\in\mathcal{P}\) and write \(g(I):=x(p,I)\). Demand is \(C^2\) in \((p,I)\).

\emph{(1 \(\Rightarrow\) 2).}
If \(\succeq\) is homothetic, then for any \(t>0\) the scaled budget set satisfies \(\{x: p\cdot x\le tI\}=t\{x: p\cdot x\le I\}\), and \(x(p,tI)=t\,x(p,I)\).
Fix \(I>0\) and define \(h(t):=x(p,tI)\). Then \(h(t)=t\,h(1)\).
By the chain rule, \(h'(t)=\partial_I x(p,tI)\cdot I\) and \(h''(t)=\partial_I^2 x(p,tI)\cdot I^2\).
Differentiating \(h(t)=t\,h(1)\) gives \(h'(t)=h(1)\) and \(h''(t)=0\).
Evaluating at \(t=1\) yields \(\partial_I^2 x(p,I)=0\) for all \(I>0\). Continuity extends this to \(I=0\) if \(0\in\mathcal{I}\).
As \(p\) was arbitrary, \(\partial_I^2 x\equiv 0\) on \(\mathcal{P}\times\mathcal{I}\).

\emph{(2 \(\Rightarrow\) 1).}
Assume \(\partial_I^2 x\equiv 0\). For each fixed \(p\), \(I\mapsto x(p,I)\) is affine, so there exist \(a(p),b(p)\in\mathbb{R}^\ell\) with
\[
x(p,I)=I\,a(p)+b(p)\quad \text{for all }(p,I).
\]
At income \(I=0\), the only affordable bundle under strictly positive prices is \(0\), and strong monotonicity implies \(x(p,0)=0\). Hence \(b(p)=0\) and \(x(p,I)=I\,a(p)\).
Therefore \(x(p,tI)=t\,x(p,I)\) for all \(t>0\), i.e., Marshallian demand scales linearly with income at fixed \(p\).
Consider the revealed preference induced by \(x(\cdot,\cdot)\). If \(x(p,I)\) is (directly) revealed preferred to some affordable \(y\) at \((p,I)\), then at \((p,tI)\) the feasible set is \(t\) times larger and the chosen bundle is \(x(p,tI)=t\,x(p,I)\), while \(t\,y\) is feasible. Thus the revealed preference respects radial scaling. By standard rationalisation results on compact domains, there exists a continuous, monotone, homothetic preference that rationalises \(x\). Hence \(\succeq\) is homothetic.

Uniform continuity of \(\mathcal{R}^{\mathrm{hom}}\)
follows from Proposition~\ref{prop:UC-derivatives}. For completeness, for any \(x,y\) we have
\[
\|\mathcal{R}^{\mathrm{hom}}(x)-\mathcal{R}^{\mathrm{hom}}(y)\|_{L^1_\mu}
=\int_{\mathcal{P}\times\mathcal{I}}\!\big\|\partial_I^2(x-y)(p,I)\big\|\,\mathrm{d}\mu(p,I)
\le \|\partial_I^2(x-y)\|_{L^\infty}
\le \|x-y\|_{C^{2,\infty}},
\]
so \(\mathcal{R}^{\mathrm{hom}}:(\mathcal{G},\|\cdot\|_{C^{2,\infty}})\to L^1_\mu\) is a bounded linear map (norm \(\le 1\)), hence uniformly continuous.

\subsection{Proof of Proposition \ref{prop:weak-sep-matrix}}

Throughout fix $(p,I)\in\mathcal{P}\times\mathcal{I}$. For $i\in G_1$ and $j\in G_2$ recall
\[
\mathcal{S}_{ij}(x_{\succeq}) \;=\; \frac{\partial x_{\succeq,i}}{\partial p_j}
\;+\; x_{\succeq,j}\,\frac{\partial x_{\succeq,i}}{\partial I},
\qquad
a_i \;:=\; \partial_I x_{\succeq,i},
\qquad
b_j \;:=\; \partial_I x_{\succeq,j},
\]
and let $\mathcal{S}^{12}(x_{\succeq})$ denote the between–group Slutsky block with entries
$\mathcal{S}_{ij}(x_{\succeq})$ for $i\in G_1$, $j\in G_2$. We will suppress the explicit
$(p,I)$–dependence to lighten notation.

\smallskip
\noindent\emph{(Necessity).}
Assume $\succeq$ is weakly separable across $(G_1,G_2)$, that is, there exist subutility
indices $v_1:\mathbb{R}^{|G_1|}_+\to\mathbb{R}$, $v_2:\mathbb{R}^{|G_2|}_+\to\mathbb{R}$ and an
aggregator $w:\mathbb{R}^2\to\mathbb{R}$ such that
\[
u(x) \;=\; w\big(v_1(x_{G_1}),\,v_2(x_{G_2})\big)
\quad\text{represents } \succeq.
\]
Under our $C^2$ and interiority assumptions, the Hicksian compensated problem is well
defined and the Slutsky matrix is the Jacobian of Hicksian demand:
$\mathcal{S}=\big(\partial h/\partial p\big)^\top$.
By the classical Goldman–Uzawa factorisation for weak separability, the between–group
block of the Hicksian substitution matrix has rank one, with entries proportional to the
cross–group income effects,\footnote{A clean way to see this is to (i) define the group
expenditure functions $e_1(p_{G_1},y_1):=\min\{p_{G_1}\!\cdot x_{G_1}:v_1(x_{G_1})\ge y_1\}$
and $e_2(p_{G_2},y_2)$; (ii) write the overall expenditure function as
$e(p,u)=\min_{y_1,y_2}\{e_1(p_{G_1},y_1)+e_2(p_{G_2},y_2):w(y_1,y_2)\ge u\}$; (iii) apply
the envelope theorem and the FOCs for the $(y_1,y_2)$–minimisation to express the
cross–group compensated derivatives $\partial h_i/\partial p_j$ as a common scalar (depending
on $p,I$) times a product of group–specific terms. Identifying those group–specific terms
with $a_i$ and $b_j$ (via Roy/Slutsky identities) yields the display below.}
hence there exists a scalar function $\kappa=\kappa(p,I)$ such that for all $i\in G_1$,
$j\in G_2$,
\begin{equation}\label{eq:GU-factor}
\mathcal{S}_{ij}(x_{\succeq}) \;=\; \kappa\,a_i\,b_j.
\end{equation}
In matrix form, \eqref{eq:GU-factor} reads
$\mathcal{S}^{12}(x_{\succeq})=\kappa\,a\,b^{\!\top}$, which is \eqref{eq:rankone-matrix}.
Moreover, fixing arbitrary $i,i'\in G_1$ and $j,j'\in G_2$ and using \eqref{eq:GU-factor},
\[
\mathcal{S}_{ij}\,a_{i'}\,b_{j'} \;-\; \mathcal{S}_{i'j'}\,a_i\,b_j
\;=\; \kappa\,a_i b_j a_{i'} b_{j'} \;-\; \kappa\,a_{i'} b_{j'} a_i b_j \;=\; 0,
\]
so each component collected by $\mathcal{R}^{\mathrm{sep}}(x_{\succeq})$ vanishes. This proves
the necessity claims.

We record the elementary equivalence we will use in sufficiency.

\begin{lem}
\label{lem:lemProofSep}
     Let $A\in\mathbb{R}^{m\times n}$, $u\in\mathbb{R}^m$, $v\in\mathbb{R}^n$.
Suppose $u$ and $v$ are not identically zero and that for all row indices
$r,r'\in\{1,\dots,m\}$ and column indices $c,c'\in\{1,\dots,n\}$,
\begin{equation}\label{eq:minor-identity}
A_{rc}\,u_{r'}\,v_{c'} \;-\; A_{r'c'}\,u_r\,v_c \;=\; 0.
\end{equation}
If there exist $r_0,c_0$ with $u_{r_0}\neq 0$ and $v_{c_0}\neq 0$, then setting
$\lambda:=A_{r_0c_0}/(u_{r_0}v_{c_0})$ yields $A=\lambda\,u\,v^{\!\top}$.
\end{lem}

\begin{proof}
    Fix any $r,c$. Plugging $(r',c')=(r_0,c_0)$ into \eqref{eq:minor-identity} gives
\[
A_{rc}\,u_{r_0}\,v_{c_0} \;=\; A_{r_0c_0}\,u_r\,v_c.
\]
Since $u_{r_0}v_{c_0}\neq 0$, we may divide both sides to obtain
$A_{rc}=\lambda\,u_r v_c$ with $\lambda=A_{r_0c_0}/(u_{r_0}v_{c_0})$. This holds for
all $r,c$, hence $A=\lambda\,u\,v^{\!\top}$.
\end{proof}

\smallskip
\noindent\emph{(Sufficiency).}
Assume the two–sided nondegeneracy condition: for the given $(p,I)$ there exist
$i_0\in G_1$ and $j_0\in G_2$ with $a_{i_0}\neq 0$ and $b_{j_0}\neq 0$. Assume further
that $\mathcal{R}^{\mathrm{sep}}(x_{\succeq})\equiv 0$, i.e., for all $i,i'\in G_1$ and
$j,j'\in G_2$,
\begin{equation}\label{eq:vanish}
\mathcal{S}_{ij}(x_{\succeq})\,a_{i'}\,b_{j'} \;-\;
\mathcal{S}_{i'j'}(x_{\succeq})\,a_i\,b_j \;=\; 0.
\end{equation}
Apply Lemma~\ref{lem:lemProofSep} with $A=\mathcal{S}^{12}(x_{\succeq})$, $u=a$, $v=b$, and
$(r_0,c_0)=(i_0,j_0)$. The hypothesis \eqref{eq:minor-identity} is exactly
\eqref{eq:vanish}. By nondegeneracy, $a_{i_0}b_{j_0}\neq 0$, so Lemma~\ref{lem:lemProofSep} yields a scalar
\[
\kappa \;:=\; \frac{\mathcal{S}_{i_0 j_0}(x_{\succeq})}{a_{i_0}\,b_{j_0}}
\quad\text{such that}\quad
\mathcal{S}^{12}(x_{\succeq}) \;=\; \kappa\,a\,b^{\!\top}.
\]
Equivalently, for all $i\in G_1$, $j\in G_2$,
$\mathcal{S}_{ij}(x_{\succeq})=\kappa\,a_i b_j$, which is \eqref{eq:rankone-matrix}.

By \citet{goldman1964note}, the rank–one factorisation of the between–group block
$\mathcal{S}^{12}(x_{\succeq})=\kappa\,a\,b^{\!\top}$ for all $(p,I)$ is equivalent to weak
separability across $(G_1,G_2)$. This completes the sufficiency part.

Finally, $\mathcal{R}^{\mathrm{sep}}$ is built from finite sums of products of terms of the form
$\mathsf{D}_p x_{\succeq}$ and $\partial_I x_{\succeq}$. By
Proposition~\ref{prop:UC-derivatives}, the maps
$x\mapsto \mathsf{D}_p x$ and $x\mapsto \partial_I x$ are bounded linear maps into
$L^1_\mu$, hence uniformly continuous. On a compact domain, pointwise products and
finite sums of uniformly continuous maps are uniformly continuous into $L^1_\mu$, so
$\mathcal{R}^{\mathrm{sep}}$ is uniformly continuous into $L^1_\mu$ as claimed.

\subsection{Proof of Proposition \ref{prop:complements}}

Let $(\cdot)^+:\mathbb{R}\to\mathbb{R}_{\ge 0}$ denote the positive–part map,
$t^+:=\max\{0,t\}$. We will use the elementary equivalence
\begin{equation}\label{eq:pospart-zero}
t^+=0 \quad \Longleftrightarrow \quad t\le 0,
\end{equation}
valid for every $t\in\mathbb{R}$.

\smallskip
\noindent\emph{(Gross complementarity).}
Fix distinct $i\neq j$. By definition, goods $i$ and $j$ are \emph{gross complements}
iff
\begin{equation}\label{eq:gross-comp-def}
\frac{\partial x_{\succeq,i}(p,I)}{\partial p_j}\;\le\;0
\qquad\text{for all }(p,I)\in\mathcal{P}\times\mathcal{I}.
\end{equation}
The proposed functional restriction is
\[
\mathcal{R}^{\mathrm{grossC}}_{ij}(x_{\succeq})(p,I)
:=\Big(\tfrac{\partial x_{\succeq,i}(p,I)}{\partial p_j}\Big)^+
= \max\!\Big\{0,\;\tfrac{\partial x_{\succeq,i}(p,I)}{\partial p_j}\Big\}.
\]

\emph{($\Rightarrow$)} If \eqref{eq:gross-comp-def} holds, then for each fixed $(p,I)$ we have
$\partial x_{\succeq,i}/\partial p_j\le 0$, hence, by \eqref{eq:pospart-zero},
$\big(\partial x_{\succeq,i}/\partial p_j\big)^+=0$. Therefore
$\mathcal{R}^{\mathrm{grossC}}_{ij}(x_{\succeq})(p,I)=0$ for all $(p,I)$, i.e.
$\mathcal{R}^{\mathrm{grossC}}_{ij}(x_{\succeq})\equiv 0$.

\emph{($\Leftarrow$)} Conversely, if $\mathcal{R}^{\mathrm{grossC}}_{ij}(x_{\succeq})\equiv 0$, then for each
$(p,I)$ we have $\big(\partial x_{\succeq,i}/\partial p_j\big)^+=0$. By
\eqref{eq:pospart-zero}, this implies $\partial x_{\succeq,i}/\partial p_j\le 0$ pointwise, proving
\eqref{eq:gross-comp-def}. Hence $i$ and $j$ are gross complements.

Combining the two implications yields
\[
i,j \text{ are gross complements}
\quad\Longleftrightarrow\quad
\mathcal{R}^{\mathrm{grossC}}_{ij}(x_{\succeq})\equiv 0.
\]

\smallskip
\noindent\emph{(Analogous forms).}
\begin{itemize}
\item \emph{Gross substitutability.} Define
$\mathcal{R}^{\mathrm{grossS}}_{ij}(x_{\succeq})(p,I):=
\big(-\partial x_{\succeq,i}(p,I)/\partial p_j\big)^+$. Repeating the previous argument with
$t=-\partial x_{\succeq,i}/\partial p_j$ shows
\[
i,j \text{ are gross substitutes}
\quad\Longleftrightarrow\quad
\mathcal{R}^{\mathrm{grossS}}_{ij}(x_{\succeq})\equiv 0.
\]

\item \emph{Net complementarity/substitutability.} Let $h(p,u)$ denote Hicksian demand and
$\mathcal{S}_{ij}(x_{\succeq})(p,I):=\partial x_{\succeq,i}/\partial p_j
+ x_{\succeq,j}\,\partial x_{\succeq,i}/\partial I$ the Slutsky term. The
Hicks–Allen definition says that $i$ is a net complement (substitute) of $j$ iff
$\partial h_i(p,u)/\partial p_j\le 0$ (resp. $\ge 0$) for all $(p,u)$.
By the Slutsky decomposition, $\partial h_i/\partial p_j=\mathcal{S}_{ij}(x_{\succeq})$ evaluated at the Marshallian pair $(p,I)$ corresponding to $(p,u)$. Hence the two sign conditions are equivalent to
$\mathcal{S}_{ij}(x_{\succeq})(p,I)\le 0$ and $\mathcal{S}_{ij}(x_{\succeq})(p,I)\ge 0$, respectively, for all $(p,I)$.
Define
\[
\mathcal{R}^{\mathrm{netC}}_{ij}(x_{\succeq})(p,I):=\big(\mathcal{S}_{ij}(x_{\succeq})(p,I)\big)^+,
\qquad
\mathcal{R}^{\mathrm{netS}}_{ij}(x_{\succeq})(p,I):=\big(-\mathcal{S}_{ij}(x_{\succeq})(p,I)\big)^+.
\]
Applying \eqref{eq:pospart-zero} pointwise yields
\[
\begin{aligned}
i,j \text{ are net complements}
&\quad\Longleftrightarrow\quad
\mathcal{R}^{\mathrm{netC}}_{ij}(x_{\succeq})\equiv 0,\\
i,j \text{ are net substitutes}
&\quad\Longleftrightarrow\quad
\mathcal{R}^{\mathrm{netS}}_{ij}(x_{\succeq})\equiv 0.
\end{aligned}
\]
\end{itemize}

\smallskip
Finally, note that on compact $\mathcal{P}\times\mathcal{I}$, the map $t\mapsto t^+$ is 1–Lipschitz, so composing it with
$(p,I)\mapsto \partial x_{\succeq,i}/\partial p_j$ or $(p,I)\mapsto \mathcal{S}_{ij}(x_{\succeq})$ preserves
uniform continuity. Since these derivatives are uniformly continuous into $L^1_\mu$ by
Proposition~\ref{prop:UC-derivatives}, the same holds for
$\mathcal{R}^{\mathrm{grossC}}_{ij}$, $\mathcal{R}^{\mathrm{grossS}}_{ij}$,
$\mathcal{R}^{\mathrm{netC}}_{ij}$, and $\mathcal{R}^{\mathrm{netS}}_{ij}$. This completes the proof.

\subsection{Proof of Proposition \ref{prop:bet-ffr-necessity}}

Throughout, let $u:\mathbb{R}_{>0}\times [0,1]\to\mathbb{R}$ be the Bernoulli index in the 
betweenness representation. For a function of two variables $u(x,v)$ we use the standard 
notation:
\[
u_1(x,v)=\frac{\partial u}{\partial x}(x,v), 
\qquad 
u_2(x,v)=\frac{\partial u}{\partial v}(x,v),
\]
and
\[
u_{11}(x,v)=\frac{\partial^2 u}{\partial x^2}(x,v),
\quad 
u_{12}(x,v)=\frac{\partial^2 u}{\partial x\partial v}(x,v),
\quad 
u_{22}(x,v)=\frac{\partial^2 u}{\partial v^2}(x,v),
\]
whenever these derivatives exist. 

Fix $(\mathbf{p},I,\boldsymbol{\pi})$ and write 
$\mathbf{x}=x_{\succeq}(\mathbf{p},I,\boldsymbol{\pi})\in\mathbb{R}^S_{>0}$ for the interior maximiser of
$V(\mathbf{x};\boldsymbol{\pi})$ subject to $\mathbf{p}\cdot\mathbf{x}\le I$.
Monotonicity of $u$ implies the budget binds, and strict quasi–concavity yields uniqueness.

Define $F(\mathbf{x},V;\boldsymbol{\pi}):=\sum_{t=1}^S \pi_t\,u(x_t,V)-V$. By \eqref{eq:betweenness-utility}, $F(\mathbf{x},V(\mathbf{x};\boldsymbol{\pi});\boldsymbol{\pi})=0$. Differentiate this identity with respect to $x_s$. Since only the term $t=s$ depends directly on $x_s$, we obtain
\[
\pi_s\,u_1(x_s,V) \;+\; \Big(\sum_{t=1}^S \pi_t\,u_2(x_t,V)\Big)\,V_{x_s} \;-\; V_{x_s} \;=\; 0 .
\]
Hence
\begin{equation}\label{eq:Vxs}
V_{x_s}
=\frac{\pi_s\,u_1(x_s,V)}{\,1-\sum_{t=1}^S \pi_t\,u_2(x_t,V)\,}\,.
\end{equation}
At an interior optimum the first–order conditions imply $V_{x_s}=\lambda p_s$ for some Lagrange multiplier $\lambda>0$, so $V_{x_s}$ is finite and strictly positive. Therefore the denominator in \eqref{eq:Vxs} is nonzero at the optimum.

From $V_{x_s}=\lambda p_s$ and \eqref{eq:Vxs}, taking the ratio $s$ to $1$ cancels the common denominator and yields
\begin{equation}\label{eq:k-ratio}
\frac{\pi_s\,u_1(x_s,V)}{\pi_1\,u_1(x_1,V)}=\frac{p_s}{p_1}
\quad\Longleftrightarrow\quad
\frac{u_1(x_1,V)}{u_1(x_s,V)}=\frac{\pi_s}{\pi_1}\frac{p_1}{p_s}=:k_s .
\end{equation}
In particular, for $s=2$ we have
\begin{equation}\label{eq:12-ratio}
\frac{u_1(x_1,V)}{u_1(x_2,V)}=k_2.
\end{equation}
By \citet[p.~314]{dekel1986axiomatic}, the fixed point $V(\mathbf{x};\boldsymbol{\pi})$ is unique for each $(\mathbf{x},\boldsymbol{\pi})$. Hence the unique value $V=V(\mathbf{x};\boldsymbol{\pi})$ at the optimum satisfies \eqref{eq:12-ratio}. Next, define
\[
\phi(v;x_1,x_2,k_2):=u_1(x_1,v)-k_2\,u_1(x_2,v).
\]
At the realised optimum, $v=V(\mathbf{x};\boldsymbol{\pi})$ solves $\phi(v;x_1,x_2,k_2)=0$ by \eqref{eq:12-ratio}. 
Assume the following local regularity holds at the realised point:
\begin{equation}\label{eq:R}
\partial_v\phi\big(V(\mathbf{x};\boldsymbol{\pi});x_1,x_2,k_2\big)
=\;u_{12}(x_1,V)-k_2\,u_{12}(x_2,V)\;\neq\;0.
\end{equation}
This is a mild pointwise condition that ensures the root of $\phi(\cdot;x_1,x_2,k_2)$ is locally unique as a function of $(x_1,x_2,k_2)$. Under \eqref{eq:R}, the Implicit Function Theorem gives a $C^1$ map
\[
V^\ast=V^\ast(x_1,x_2,k_2)
\quad\text{with}\quad
\phi\big(V^\ast(x_1,x_2,k_2);x_1,x_2,k_2\big)=0
\]
in a neighbourhood of the realised $(x_1,x_2,k_2)$. In particular, at the optimum $V^\ast(x_1,x_2,k_2)=V(\mathbf{x};\boldsymbol{\pi})$. Using \eqref{eq:k-ratio} with $s\ge 3$ and substituting $V^\ast$,
\begin{equation}
\label{eq: prof k_2/k_s}
    \frac{u_1(x_s,V^\ast)}{u_1(x_2,V^\ast)}=\frac{k_2}{k_s}.
\end{equation}
For each fixed $v$, the map $x\mapsto u_1(x,v)$ is strictly decreasing and continuous because $u_1>0$ and $u_{11}<0$. Hence it admits a continuous inverse in its first argument, denoted $h(\cdot\,;v)$. Invert \eqref{eq: prof k_2/k_s} in $x_s$ to obtain
\[
x_s
= h\!\left(\frac{k_2}{k_s}\,u_1(x_2,V^\ast(x_1,x_2,k_2))\ ;\ V^\ast(x_1,x_2,k_2)\right)
=: f(x_1,x_2,k_2,k_s) .
\]
Monotonicity in $k_s$ follows because $k_s\mapsto (k_2/k_s)$ is strictly decreasing and $r\mapsto h(r;v)$ is strictly decreasing, so the composition is strictly increasing in $k_s$. 
For the normalisation, if $x_1=x_2=x$ and $k_s=1$, then $u_1(x_s,V^\ast)=u_1(x,V^\ast)$ and therefore $x_s=x$, which gives $f(x,x,k_2,1)=x$. Since the construction holds at each $(\mathbf{p},I,\boldsymbol{\pi})$, we have
\[
x_{\succeq,s}(\mathbf{p},I,\boldsymbol{\pi})
= f\big(x_{\succeq,1}(\mathbf{p},I,\boldsymbol{\pi}),\,x_{\succeq,2}(\mathbf{p},I,\boldsymbol{\pi}),\,k_2(\mathbf{p},\boldsymbol{\pi}),\,k_s(\mathbf{p},\boldsymbol{\pi})\big)
\]
with $f$ strictly increasing in $k_s$ and satisfying $f(x,x,k_2,1)=x$.

\subsection{Proof of Proposition~\ref{prop:Rbet}}

Fix $s\ge 3$. By Proposition~\ref{prop:bet-ffr-necessity}, locally $x_{\succeq,s}=f\big(x_{\succeq,1},\,x_{\succeq,2},\,k_2,\,k_s\big)$, with $f$ as constructed in the proof of the result. Consider the set of admissible perturbations of $(\mathbf{p},I)$ that keep $k_2$ fixed to first order. Formally, let $\Psi:\mathbb{R}^m\to\mathbb{R}^{S+1}$ be a $C^1$ local parametrisation of this subspace, with coordinates $\theta\in\mathbb{R}^m$, so that $(\mathbf{p},I)=(\mathbf{p}(\theta),I(\theta))$ and $\dv{x}k_2(\mathbf{p}(\theta),\boldsymbol{\pi})\big|_{\theta=0}=0 $ in every coordinate direction. Define 
\[
H_s(\theta):=\big(x_{\succeq,1}(\mathbf{p}(\theta),I(\theta),\boldsymbol{\pi}),\ 
x_{\succeq,2}(\mathbf{p}(\theta),I(\theta),\boldsymbol{\pi}),\ 
k_s(\mathbf{p}(\theta),\boldsymbol{\pi}),\ 
x_{\succeq,s}(\mathbf{p}(\theta),I(\theta),\boldsymbol{\pi})\big).
\]
Let $J_s$ be the $4\times m$ Jacobian of $H_s$ with respect to $\theta$ at $\theta=0$.
By the chain rule applied to $x_{\succeq,s}=f(x_{\succeq,1},x_{\succeq,2},k_2,k_s)$ and using $\dd k_2=0$ along $\Psi$,
\[
\dd x_{\succeq,s}
= f_{x_1}\,\dd x_{\succeq,1} \;+\; f_{x_2}\,\dd x_{\succeq,2} \;+\; f_{k_s}\,\dd k_s .
\]
Therefore the fourth row of $J_s$ is a linear combination of the first three rows with coefficients $(f_{x_1},f_{x_2},f_{k_s})$. It follows that $\rank(J_s)\le 3$, so every $4\times 4$ minor of $J_s$ vanishes. Defining
\[
\mathcal{R}^{\mathrm{bet}}_{s}(x_{\succeq})(\mathbf{p},I,\boldsymbol{\pi})
:=\sum_{M\in\mathscr{M}_{4\times 4}(J_s)} (\det M)^2
\]
gives $\mathcal{R}^{\mathrm{bet}}_{s}\equiv 0$ as a necessary condition for betweenness rationalisability when the Jacobian is evaluated along directions with $\dd k_2=0$.

%

\section{Simulations} \label{simulation:dgp}

In this section, we first present the pertubation we used in our simulations and prove its properties. As stated in Section \ref{ss: finite-sample}, our results apply with mean-zero measurement error. As such, we include simulations and power curves for this extension.

\subsection{Perturbation of Cobb--Douglas Demand}

We consider demands of the form $y = x + \varepsilon f$, where $f$ is a perturbation normalized so that the Frobenius norm of the induced skew--symmetric Slutsky component satisfies
\[
\|\operatorname{Skew}(S^{f})\|_F = 1.
\]
Thus $\varepsilon$ indexes the distance from Slutsky symmetry, while different choices of $f$ allow us to explore different patterns of violation. Let
\[
f^A(p,I) = 
\begin{pmatrix}
g_{12}(p,I) \\[1mm]
- \frac{p_1}{p_2} g_{12}(p,I) \\[1mm]
0 \\[1mm]
0
\end{pmatrix}, 
\qquad
f^B(p,I) = 
\begin{pmatrix}
0 \\[1mm]
0 \\[1mm]
g_{34}(p,I) \\[1mm]
- \frac{p_3}{p_4} g_{34}(p,I)
\end{pmatrix},
\]
with
\[
g_{12}(p,I)=\Big(\frac{I}{p_1}\Big)^2 \frac{p_2}{p_1+p_2}, 
\qquad
g_{34}(p,I)=\Big(\frac{I}{p_3}\Big)^2 \frac{p_4}{p_3+p_4}.
\]

\noindent
In the simulations presented in the main text, we considered the perturbation $f = f^A + f^B$. As a robustness check, we also considered the perturbation $f = \alpha f^A + (1-\alpha)f^B$ with $\alpha = 0.1$. In what follows, we prove the properties for the perturbation and detail the normalization.

\paragraph{Homogeneity of degree zero.}
Each block is homogeneous of degree zero jointly in $(p,I)$. Indeed, for any $\lambda>0$, we have $g_{12}(\lambda p,\lambda I) = g_{12}(p,I)$ and $g_{34}(\lambda p,\lambda I) = g_{34}(p,I)$ such that $f^A(\lambda p,\lambda I) = f^A(p,I)$, $f^B(\lambda p,\lambda I) = f^B(p,I)$, and $f(\lambda p,\lambda I) = f(p,I)$.

\paragraph{Budget neutrality.}
Each block is budget neutral:
\[
p\cdot f^A = p_1 g_{12} - p_1 g_{12} = 0, \qquad
p\cdot f^B = p_3 g_{34} - p_3 g_{34} = 0.
\]
Hence, for $f = f^A + f^B$ and $f = \alpha f^A + (1-\alpha)f^B$, we have $p\cdot f = 0$.

\paragraph{Violation of Slutsky symmetry.}
Consider the perturbed demand
\[
y(p,I) = x(p,I) + f(p,I),
\]
where $x(p,I)$ are the base Cobb-Douglas demands and $f(p,I)$ is the perturbation designed to violate integrability. The corresponding Slutsky matrix is
\[
S(p,I) = \mathsf{D}_p y(p,I) + y(p,I) \, (\partial_I y(p,I))^\top,
\]
where $\mathsf{D}_p y$ and $\partial_I y$ denote the derivatives of $y$ with respect to prices and income, respectively. For the first perturbation, the induced asymmetries in the Slutsky matrix are
\[
S^A_{12} - S^A_{21} = \frac{I^2(\alpha_1+\alpha_2-1)}{p_1^2(p_1+p_2)}, 
\qquad
S^B_{34} - S^B_{43} = \frac{I^2(\alpha_3+\alpha_4-1)}{p_3^2(p_3+p_4)}.
\]
All other off-diagonal entries of the skew part vanish. Therefore, the Slutsky matrix is generically asymmetric, and the asymmetry is split between $(1,2)$ and $(3,4)$. For the second perturbation, the argument is identical but with the scale $\alpha$ and $1-\alpha$ on the asymmetry $(1,2)$ and $(3,4)$, respectively.

\paragraph{Normalization to control the magnitude of deviation.} 
Each perturbation $f$ is scaled so that the Frobenius norm of the induced
skew--symmetric Slutsky component equals one:
\[
\hat{f} = \frac{f}{\|\operatorname{Skew}(S^f)\|_F},
\]

\noindent
where $S^f$ denote the Slutsky matrix for demand with perturbation $f$ and 
\[
\operatorname{Skew}(S^f) = \frac{1}{2}\big(S^f - (S^f)^\top\big)
\]
is its skew--symmetric part. After this rescaling, we replace $f$ by $\tilde f$ in the perturbed demand:
\[
y = x + \varepsilon \tilde f.
\]

\noindent
After this step, $\varepsilon$ measures the total magnitude of the violation from rationality in the same way across perturbations.

\subsection{Additional Regressions}

This section includes a log-linear regression and a log-linear regression with reciprocal predictor $(1/\varepsilon)$ to complement Table \ref{tab:combined_power}. The results for the log-linear regression are presented in Table \ref{tab:combined_power1}

\begin{table}[h!]
\centering
\caption{Log-linear regression estimates from power curves}
\label{tab:combined_power1}
\begin{tabular}{lcccc}
\toprule
Parameter & SUM & H & WS & EU \\
\midrule
$\beta_0$ & 4.2168 & 2.4133 & 3.2303 & 2.6499 \\
$\beta_1$ & -4.2316 & -0.0870 & -3.0240 & -4.8806 \\
\bottomrule
\end{tabular}
\end{table}

Table \ref{tab:combined_power1} shows that the required sample size decreases progressively from SUM to WS, then EU, and finally HARP. The table also shows that SUM and EU have higher sensitivity of required sample sizes to deviations from rationality than H and WS. Since sample size is modeled on a log scale, EU has a steeper slope ($\beta_1$) in terms of relative changes even though its curve appears flatter in absolute sample size due to a lower baseline sample size ($\beta_0$) compared with WS. Next, the results for a log-linear regression with a reciprocal preditor are presented in Table \ref{tab:combined_power2}.
\begin{table}[h!]
\centering
\caption{Log-linear regression reciprocal estimates from power curves}
\label{tab:combined_power2}
\begin{tabular}{lcccc}
\toprule
Parameter & SUM & H & WS & EU \\
\midrule
$\beta_0$ & 3.3181 & 2.3873 & 2.5749 & 1.5717 \\
$\beta_1$ & 0.02791 & 0.001209 & 0.02107 & 0.03574 \\
\bottomrule
\end{tabular}
\end{table}

Table~\ref{tab:combined_power2} shows that SUM WS and EU require substantially larger sample sizes to detect small deviations from rationality compared to H which is relatively insensitive to the size of deviations.

\subsection{Power with Additive Measurement Error}

\begin{figure}[h!] 
    \centering
    \begin{subfigure}{0.49\textwidth}
        \centering
        \includegraphics[width=\textwidth]{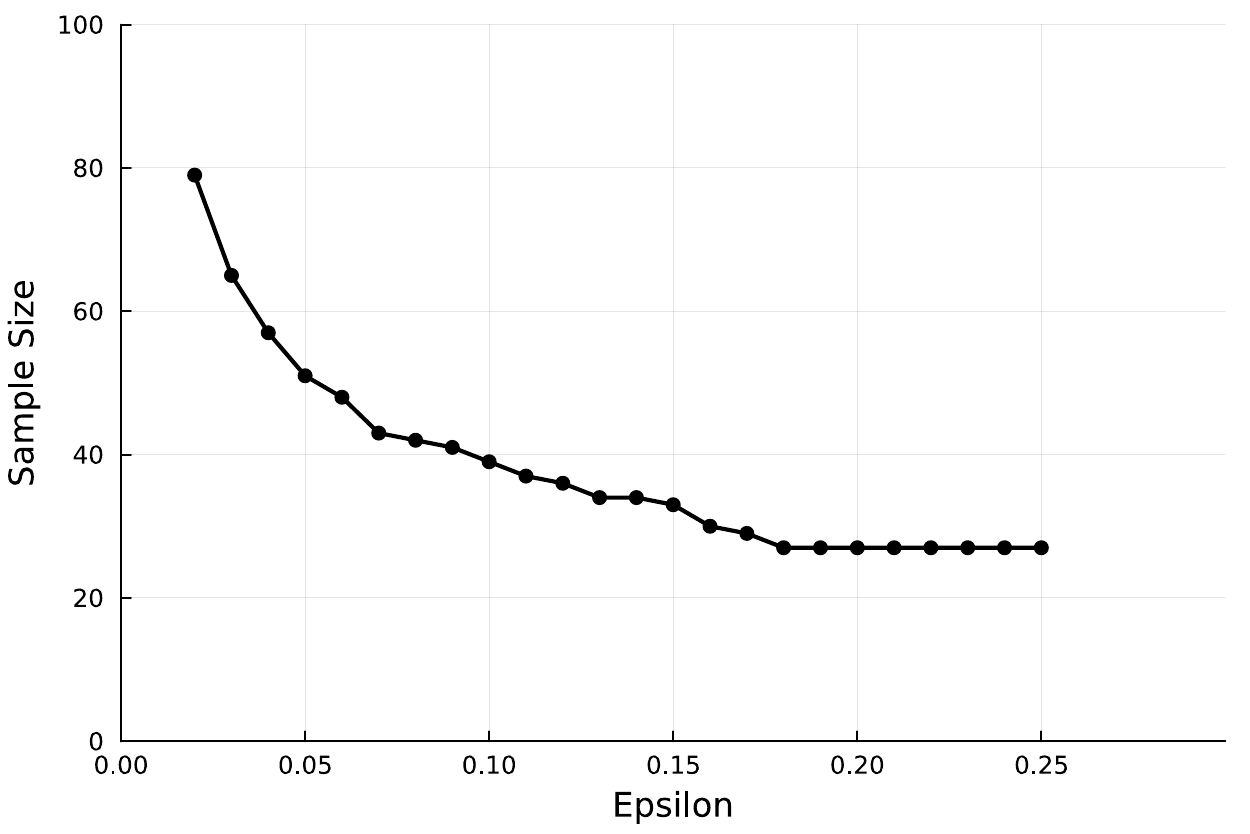}
        \caption{SUM}
        \label{fig:powerGARP}
    \end{subfigure}
    \hfill
    \begin{subfigure}{0.49\textwidth}
        \centering
        \includegraphics[width=\textwidth]{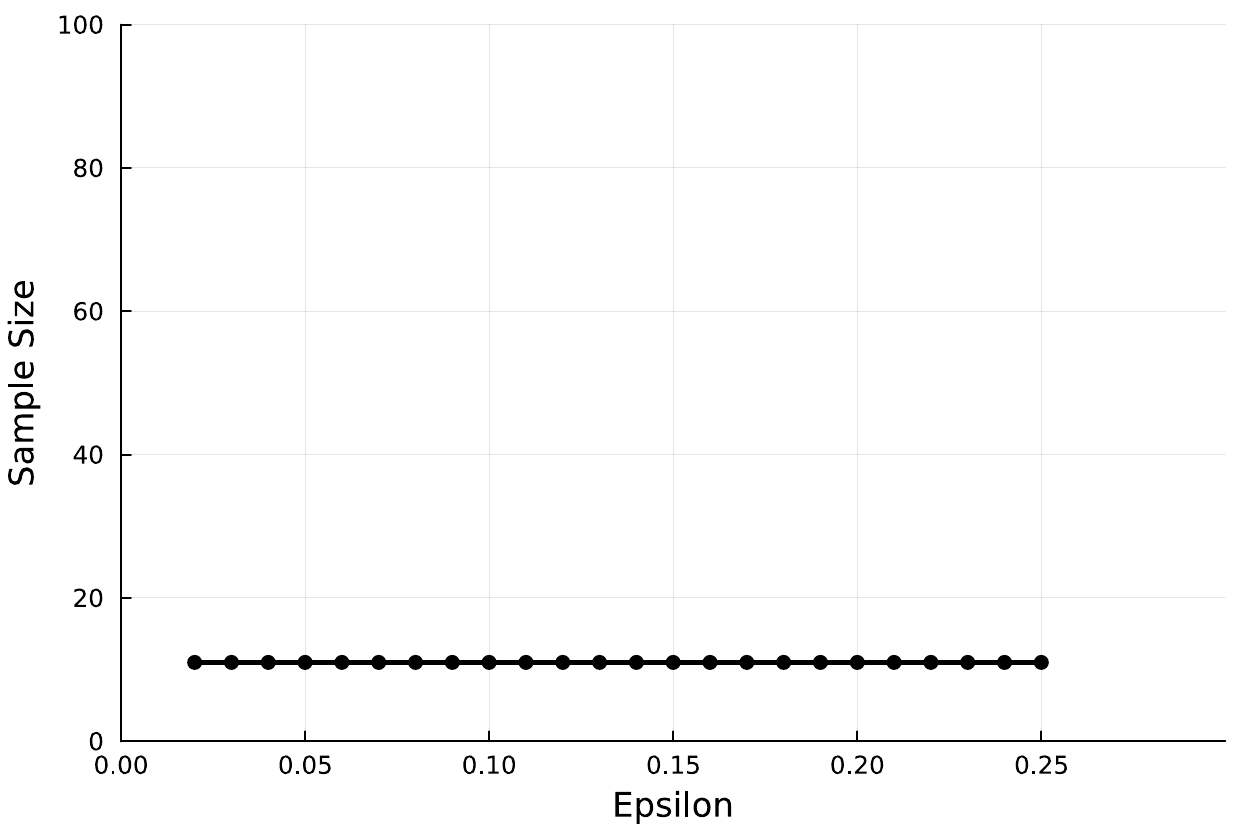}
        \caption{H}
        \label{fig:powerHARP}
    \end{subfigure}

    \vspace{1em}

    \begin{subfigure}{0.49\textwidth}
        \centering
        \includegraphics[width=\textwidth]{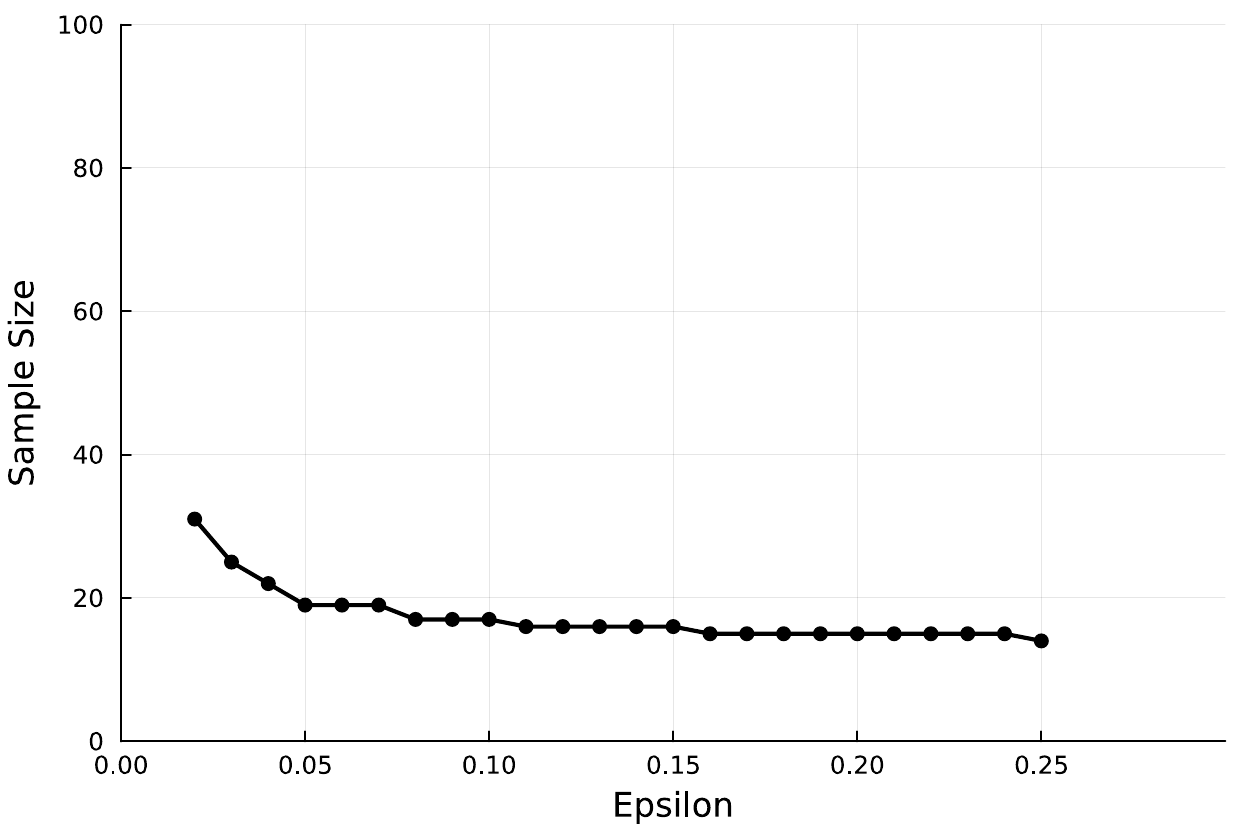}
        \caption{WS}
        \label{fig:powerWS}
    \end{subfigure}
    \hfill
    \begin{subfigure}{0.49\textwidth}
        \centering
        \includegraphics[width=\textwidth]{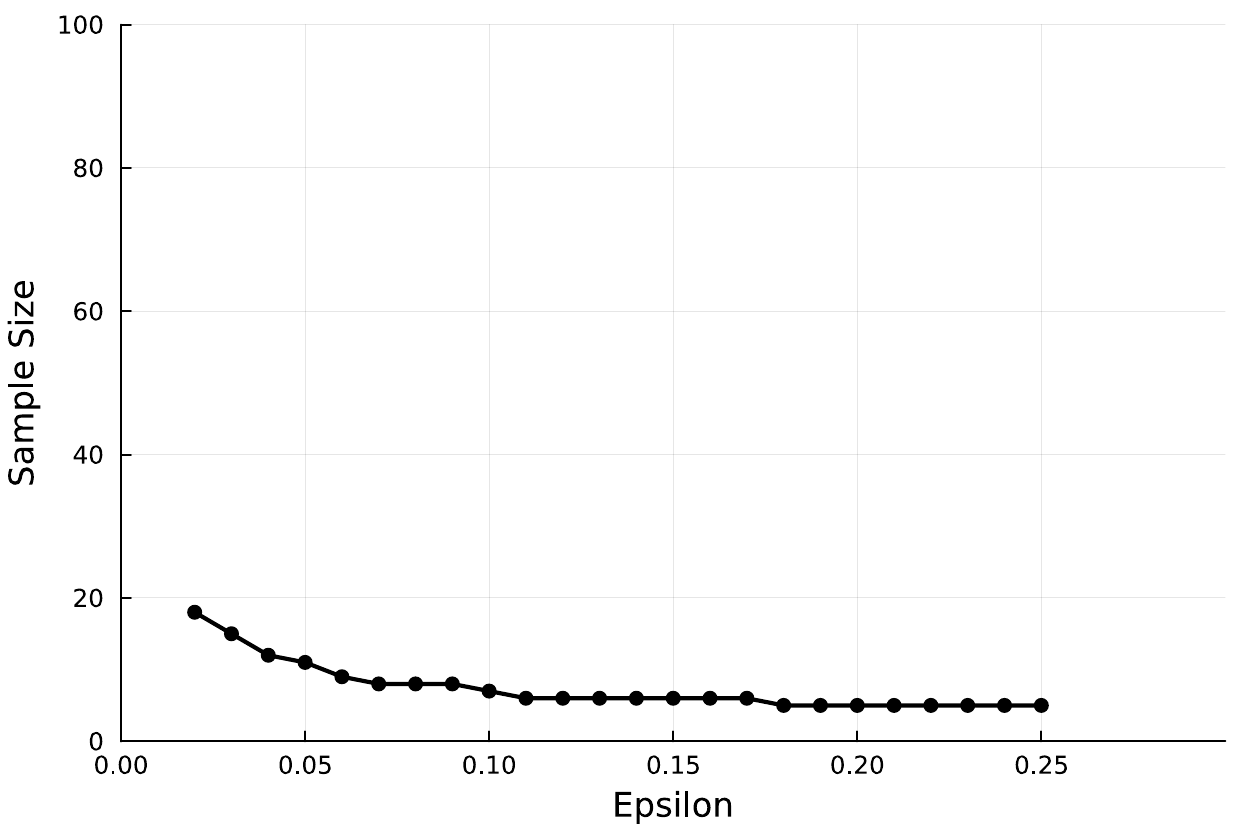}
        \caption{EU}
        \label{fig:powerEU}
    \end{subfigure}

    \caption{Power curves for SUM, H, WS, and EU.}
    \label{fig:power_all_me}
\end{figure}

To model measurement error in observed demands, we add mean-zero additive noise to each demand component:
\[
\tilde{x}_k = x_k + m_k,
\]
where the measurement error $m_k$ is drawn independently according to a truncated normal distribution with standard deviation $0.25$ and support set to $\pm50\%$ of the demand $x_k$. This support implies that the noise is heteroskedastic and proportional to the magnitude of the demand. The noisy demands are scaled back to ensure the budget constraint holds. The data generating process and the perturbation are otherwise identical to that in the main text. In particular, the perturbation is computed for the demand $x_k$. The power curves are reported in Figure \ref{fig:power_all_me}

The power curves are qualitatively similar to those in the main text, thus empirically confirming that the presence of mean-zero measurement error is inconsequential for our results.

\end{appendices}

\end{document}